%% file: NonSpinningPaper.tex
\documentclass[aps, prd, amsmath, amssymb, twocolumn, 10pt, a4paper, superscriptaddress, nofootinbib,
longbibliography]{revtex4-2}

\usepackage{textcomp}
\usepackage{amsfonts}
\usepackage{calc}
\usepackage{mathrsfs}
\usepackage{array}
\usepackage{xcolor}
\usepackage{bm}
\usepackage{graphicx}
\usepackage{hyperref}
\usepackage{tensor}
\usepackage{booktabs}
\usepackage{mathtools}
\usepackage[caption=false]{subfig}
\usepackage{float}
\usepackage{orcidlink}
\usepackage[normalem]{ulem}
\usepackage{placeins}
\usepackage[inline]{enumitem}

\graphicspath{{Figures/}}

\newcommand{\diff}[2]  {\frac{d #1}{d #2}}

\newcommand{\pdiff}[2]  {\frac{\partial #1}{\partial #2}}
\newcommand{\psdiff}[2]  {\frac{\partial^2 #1}{\partial #2^2}}
\newcommand{\eqn}[1] {Eq.~\eqref{eq:#1}}
\newcommand{\fig}[1] {Fig.~\ref{fig:#1}}

\newcommand{\secref}[1] {Sec.~\ref{sec:#1}}

\newcommand{\nn}{\nonumber}

\newcommand{\rstar} {r_{*}}

\DeclarePairedDelimiterX\braket[1]{\langle}{\rangle}{#1}
\newcommand{\abs}[1]{\left\lvert#1\right\rvert}

\newcommand{\beq}{\begin{equation}}
\newcommand{\eeq}{\end{equation}}
\newcommand{\bec}{\begin{cases}}
\newcommand{\eec}{\end{cases}}

\newcommand{\AEI}{Max Planck Institute for Gravitational Physics (Albert Einstein Institute), Am M\"uhlenberg 1, Potsdam 14476, Germany}
\newcommand{\UMD}{Department of Physics, University of Maryland, College Park, MD 20742, USA}

\newcommand{\COG}{\affiliation{Center of Gravity, Niels Bohr Institute, Blegdamsvej 17, 2100 Copenhagen, Denmark}}

\definecolor{colour1}{HTML}{0571b0} %--- Blue
\definecolor{colour2}{HTML}{92c5de} %--- Cyan
\definecolor{colour3}{HTML}{f4a582} %--- Orange
\definecolor{colour4}{HTML}{ca0020} %--- Maroon
\definecolor{colour5}{HTML}{fe4a49} %--- Red

\hypersetup{colorlinks=true, linkcolor=colour1, citecolor=colour1,
filecolor=colour1, urlcolor=colour1}

\begin{document}
\title{Inspiral-merger-ringdown waveforms with gravitational self-force results \\ within the effective-one-body formalism}

%\texttt{SEOBNR-GSF}: A nonspinningeffective-one-body model using gravitational self-force

\author{Benjamin Leather\,\orcidlink{0000-0001-6186-7271}} 
%\email{benjamin.leather@aei.mpg.de}
\affiliation{\AEI}
\author{Alessandra Buonanno\,\orcidlink{0000-0002-5433-1409}}
\affiliation{\AEI}
\affiliation{\UMD}
\author{Maarten van de Meent,\orcidlink{0000-0002-0242-2464}}
\COG\affiliation{\AEI}
\date{\today}

%------------------------------------------------
%	Abstract
%------------------------------------------------

\begin{abstract}
Gravitational self-force (GSF) theory is a strong-gravity perturbative approach to the relativistic 
two-body problem, primarily developed to model extreme-mass-ratio inspirals, where one compact object 
is significantly more massive than the companion. However, recent advancements in GSF calculations, particularly 
involving second-order self-force (2GSF) results, indicated a much broader applicability 
across a wider range of mass ratios. These developments have motivated efforts to incorporate GSF results 
into the effective-one-body (EOB) framework, where they have 
already been successfully integrated into the state-of-the-art waveform model, \texttt{SEOBNRv5}, 
employed in recent LIGO-Virgo-KAGRA (LVK) observing runs. 
In this work, we present \texttt{SEOBNR-GSF}, a nonspinning inspiral-merger-ringdown (IMR) EOB 
waveform model that introduces a GSF-informed EOB Hamiltonian as its central innovation. 
This marks the first complete IMR waveform model constructed primarily from GSF results.
We show that our model outperforms inspiral waveforms from full 2GSF calculations in the 
intermediate-to-comparable mass regime. 
Furthermore, by comparing with a post-Newtonian-informed variant, \texttt{SEOBNR-GSF-PN}, 
we demonstrate that the inclusion of numerical GSF information in the Hamiltonian leads to significant 
improvements in model fidelity.
Finally, we benchmark our model against high-accuracy, nonspinning numerical-relativity simulations from the 
Simulating eXtreme Spacetimes (\texttt{SXS}) catalogue and find that its median mismatch is comparable to that of \texttt{SEOBNRv5}, 
suggesting that this approach holds promise for further enhancing future EOB waveform models.
\end{abstract}

\maketitle

%------------------------------------------------
%	Introduction
%------------------------------------------------
\section{Introduction}
\label{sec:introduction}

The growth of gravitational-wave astronomy since the first observation of 
a binary black-hole (BH) merger in 2015 \cite{LIGOScientific:2016aoc} is indebted to accurate theoretical
descriptions of the two-body dynamics and gravitational radiation in general relativity (GR).  
The extension in frequency bandwith and diversity of astrophysical sources from current and future detectors, 
requires continual improvement in waveform models.
From a modelling perspective, this improvement rests in the symbiosis 
of numerical and analytical methods.
Numerical relativity (NR) \cite{Pretorius:2005gq,Baker:2005vv, PhysRevLett.96.111101}, the fully non-linear 
realisation of the two-body problem in GR, produces the most accurate waveforms, but remains computationally
expensive for practical waveform generation. This necessitates the development of approximate models for waveform 
production that are sufficiently efficient, yet do not comprimise on accuracy to support the science goals of 
upcoming GW observatories.

Three main perturbative frameworks underpin such approximate modelling schemes: post-Newtonian (PN) theory
\cite{Futamase:2007zz, Blanchet:2013haa, Porto:2016pyg, Schafer:2018jfw, Levi:2018nxp, Jaranowski:1997ky,
Damour:2014jta, Jaranowski:2015lha, Bernard:2015njp, Bernard:2016wrg, Damour:2016abl, Blanchet:2023sbv, Blanchet:2023bwj}, 
valid in the weak-field, small-velocity regime $v^{2}/c^{2} 
\sim G M /r c^{2} \ll 1$\footnote{Here $M = m_{1} + m_{2}$ is the total mass of system, 
where $m_{1}$ and $m_{2}$ are masses of the primary and secondary, respectively. $r$ is the
radial separation.}; post-Minkowskian (PM) theory
\cite{Westpfahl:1979gu, Westpfahl:1980mk, Bel:1981be, 
Westpfahl:1985tsl, Ledvinka:2008tk, Buonanno:2022pgc, Travaglini:2022uwo, Bjerrum-Bohr:2022blt, Kosower:2022yvp},
applicable in the weak fields whereby $G M /r c^{2} \ll 1$, but allowing relativistic velocities $v^{2}/c^{2} \leq 1$;
and gravitational self-force (GSF) theory, which captures strong-field dynamics in the small mass-ratio limit, 
$q = m_{2}/m_{1} \ll 1$.
While each of these frameworks provides crucial insights into different aspects of the two-body problem’s 
dynamics, none can, on its own, accurately model the complete evolution of a binary across the entire parameter space. 
The integration of these methods would allow for a more comprehensive and precise description of the waveform 
throughout all phases of a binary’s evolution.

For ground-based detectors, three primary methods have been developed to construct complete waveform models that
span the entire parameter space of compact binaries: NR surrogate models, phenomenological models, and effective-one-body
(EOB) models.
Surrogate models 
\cite{Blackman:2015pia, Blackman:2017dfb, Blackman:2017pcm, Varma:2018mmi, Varma:2019csw, Williams:2019vub,
Rifat:2019ltp} interpolate between NR simulations using dimensionality reduction 
techniques, achieving high accuracy but limited in parameter space coverage and waveform duration.  
Phenomenological models \cite{Pan:2007nw, Ajith:2007qp, Ajith:2009bn, Santamaria:2010yb, Hannam:2013oca, 
Husa:2015iqa, Khan:2015jqa, London:2017bcn, Khan:2018fmp, Khan:2019kot} fit closed-form expressions 
to hybrid PN/EOB-NR waveforms, enabling fast waveform generation across broad parameter ranges, 
but relying on NR and EOB for calibration.  Effective-one-body (EOB) formalism 
\cite{Buonanno:1998gg, Buonanno:2000ef, Damour:2000we, Damour:2001tu, Buonanno:2005xu, Buonanno:2006ui} 
combines insights from two-body methods to provide a comprehensive description of compact binary dynamics.
EOB resums information from two-body methods in such a way to include the exact strong-field test-particle limit.  
More precisely, EOB maps the dynamics of a binary system to that of a test body of a deformed background spacetime, 
with the deformation being parameterised by the symmetric mass ratio, $\nu$. Being a physical approach, 
it has the advantage that not only the waveforms, but multiple observables (e.g., binding energy, periastron advance, 
scattering angle) can be compared to NR, assessing its consistency and robusteness. On the other side, since 
EOB waveforms are built in time-domain, solving ordinary differential equations makes the waveform generation slower 
than when using closed-form, frequency-domain waveforms, but they are more accurate since less approximations are 
adopted. 
Currently, two main families of EOB waveform models exist: \texttt{SEOBNR} (see 
Refs.~\cite{Pompili:2023tna, Ramos-Buades:2023ehm, vandeMeent:2023ols, Khalil:2023kep, Ramos-Buades:2021adz}), 
which will be the focus of this work and, \texttt{TEOBResumS} (see 
Refs.~\cite{Nagar:2018zoe, Nagar:2019wds, Nagar:2020pcj, Gamba:2021ydi, Nagar:2023zxh}).

Effective-one-body and phenomenological models form the backbone of waveform modelling used by
the LIGO-Virgo-Kagra (LVK) Collaboration 
\cite{LIGOScientific:2007fwp, LIGOScientific:2014pky, LIGOScientific:2014qfs, PhysRevD.102.062003, 
PhysRevLett.123.231107, VIRGO:2012dcp, VIRGO:2014yos, PhysRevLett.123.231108, 10.1093/ptep/ptaa125, 
PhysRevD.88.043007, Somiya:2011np}. EOB models have traditionally relied on 
sophisticated resummations of the PN expansion.
However, the upcoming generation of GW observatories --- including next-generation (XG) ground-based detectors like the 
Einstein Telescope \cite{Punturo:2010zz} and Cosmic Explorer \cite{Reitze:2019iox, Evans:2021gyd}, 
as well as the space-based LISA mission \cite{LISA:2017pwj} --- is expected to significantly extend the
accessible parameter space and increase signal-to-noise ratios by up to two orders of magnitude \cite{Borhanian:2022czq}.
This makes the need for more accurate and robust waveform models increasingly urgent. 
Recent work, such as Ref.~\cite{Purrer:2019jcp, Dhani:2024jja}, has shown that even state-of-the-art waveform models may 
induce significant biases when applied to future LVK and XG signals, particularly in regimes with large 
mass ratios or high BH spins.

To address the need for enhanced waveform accuracy, recent efforts have focused on developing new
methods for on incorporating perturbative results beyond PN into the EOB framework.  Notably, recent work in
Ref.~\cite{Buonanno:2024byg} has integrated PM information into the \texttt{SEOBNR} framework,
resulting in a novel IMR waveform model, \texttt{SEOBNR-PM}.  This model is built upon a
PM-expanded Hamiltonian, utilizing an expansion in $G M / r c^{2}$, as opposed to the traditional
PN-expansion in $v^{2}/c^{2}$. However, PN information is also included, where necessary. 
Preliminary comparisons indicate that \texttt{SEOBNR-PM} demonstrates
promising performance relative to the current state-of-the-art \texttt{SEOBNRv5} model, even
achieving superior agreement with NR during the inspiral phase before calibration to NR.

An equally promising direction lies in the incorporation of GSF results,
which provide precise information in the strong-field, small mass-ratio regime.
GSF theory naturally complements the EOB framework, since in the test-particle limit, the EOB dynamics 
reduce to the motion in a fixed background. Furthermore, GSF calculations distinguish between conservative 
(time-symmetric) and dissipative (time-antisymmetric) components of the dynamics, which aligns well with the 
EOB formulation. Recent advances have enabled the computation of second-order GSF (2GSF) quantities, 
including gravitational wave fluxes \cite{Warburton:2021kwk} and binding energies \cite{Pound:2019lzj}, 
and the construction of the first post-adiabatic waveforms for quasicircular, nonspinning 
binaries \cite{Wardell:2021fyy}.

To date, GSF information has primarily been incorporated into EOB models to improve the radiation-reaction sector, 
including fluxes and mode amplitudes, through calibration in the test-particle limit \cite{vandeMeent:2023ols}. 
The only conservative input so far has been the ISCO shift from first-order GSF calculations \cite{Barack:2009ey}. 
Attempts to inform the EOB Hamiltonian using GSF data face significant obstacles: most notably, the appearance of 
gauge singularities at the light-ring when evolving in the standard EOB gauge \cite{Akcay:2012ea, Antonelli:2019fmq}. 
While an alternative gauge developed in Refs.~\cite{Damour:2017zjx, Antonelli:2019fmq} has been proposed to 
avoid this pathology, it has not yet been incorporated into a full IMR EOB model.

The structure of this paper is as follows. In \secref{eob_theory}, we review the necessary 
elements of the EOB theory that are critical for constructing a GSF-informed EOB Hamiltonian. 
This includes a discussion of the light-ring gauge singularity in the standard EOB gauge and the 
alternative gauge proposed in Refs.~\cite{Damour:2017zjx, Antonelli:2019fmq} to mitigate this issue. 
\secref{ps_gauge_hamiltonian} introduces a post-Schwarzschild gauge Hamiltonian with corrections up to 
4PN order, which extends the results of Antonelli \emph{et al.} \cite{Antonelli:2019fmq} by including 
higher-order terms. Following this, \secref{imr_model} examines the \texttt{SEOBNR} framework, with a focus 
on incorporating our new Hamiltonian into a full IMR waveform model. 
In \secref{calibration_with_nr}, we outline the calibration of the model against NR simulations. 
\secref{results} presents a detailed comparison of the model’s performance against \texttt{SEOBNRv5}, 
assessing the agreement of the binding energy and waveforms
relative to 2GSF (\texttt{1PAT1}) ones, and computing mismatches with NR data. 
Finally, \secref{summary_and_discussion} concludes with a summary of our 
findings and a discussion of future research directions.

\subsection*{Notation}
We shall first introduce some conventions that are used throughout. 
We use geometrized units such that $G = c = 1$ and adopt
the metric signature $(- + + +)$.  For an exemplar binary with masses $m_{1}$
and $m_{2}$, with $m_{1} \geq m_{2}$, we define the \emph{reduced mass} to
be $\mu := {m_{1}m_{2}}/{M}$, where $M := m_{1} + m_{2}$ is the total mass
of the system.  Another important quantity is the \emph{symmetric mass ratio},
$\nu := {\mu}/{M}$. 
In the binary's center-of-mass, the EOB dynamics is described by
$q_{a} := (r, \varphi)$, with canonically conjugate momenta
$p_{a} := (p_{r}, p_{\varphi})$.  In practice, we often use the mass-reduced
inverse orbital separation, $u := M/r$, and the mass-reduced momenta
$\hat{p}_{r} := p_{r}/\mu$ and $\hat{p}_{\varphi} :=
{p_{\varphi}/\mu}$.  Similarly, any Hamiltonian written with a ``hat"
is the mass-reduced version such that $\hat{H} := {H/\mu}$.

\section{Effective-one-body theory}
\label{sec:eob_theory}
In (nonspinning) EOB, the motion of the inspiralling binary system is mapped to 
that of a test mass in a deformed Schwarzschild spacetime.  The connection between the real
Hamiltonian, $H_{\rm EOB}$, and the effective Hamiltonian, $H_{\rm eff}$ is given by the energy map:
\beq
	H_{\rm EOB} = M\sqrt{1 + 2\nu \left(\frac{H_{\rm eff}}{\mu} - 1\right)}.
	\label{eq:energy_map}
\eeq
The EOB effective metric associated with this effective Hamiltonian has a line element of the form
\beq
	g^{\rm eff}_{\mu\nu}dx^{\mu}dx^{\nu} 
	= -A(r,\nu) dt^{2} + [A(r, \nu) \Bar{D}(r, \nu)]^{-1} dr^{2} + r^{2}d\Omega^{2}
	\label{eq:eff_metric_line_element}
\eeq
together with the \emph{mass-shell condition} \cite{Damour:2000we, Damour:2017zjx},
\beq
	g^{\mu\nu}_{\rm eff}p_{\mu}p_{\nu} + \mu^{2} + Q(u, \nu, p_{r}, p_{\varphi}) = 0.
	\label{eq:mass_shell_condition}
\eeq
We have introduced the non-goedesic function, $Q(u, \nu, p_{r}, p_{\varphi})$,
in order to preserve the mapping in \eqn{eff_metric_line_element} through 3PN-order 
\cite{Antonelli:2019fmq}.
The generic form of the effective Hamiltonian, $H_{\rm eff}$, is then found 
through the simple relation, $H_{\rm eff} = -p_{0}$, yielding
\begin{align}
	H^{2}_{\rm eff} &= A(u, \nu) [ \mu^{2} + A(u, \nu)\Bar{D}(r, \nu)p^{2}_{r} \nn \\
	&+ p_{\varphi}^{2}u^{2} + Q(u, \nu, p_{r}, p_{\varphi}) ].
	\label{eq:h_eff_generic}
\end{align}
In the test mass limit, whereby $\nu \rightarrow 0$, the potentials $A(u, \nu)$ and 
$\Bar{D}(u, \nu)$ reduce to
\beq
	A(u, \nu) |_{\nu \rightarrow 0} = 1 - 2u, \quad
	\Bar{D}(u, \nu) |_{\nu \rightarrow 0} = 1,
	\label{eq:potentials_test_mass_limit}
\eeq
and hence we retain the usual Schwarzschild Hamiltonian given by
\beq
	\hat{H}^{2}_{\rm S} = (1 - 2u) \left[ 1  + (1 - 2u)p^{2}_{r} + p^{2}_{\varphi}u^{2} \right].
	\label{eq:schwarzschild_hamiltonian}
\eeq
$A(u, \nu)$, $\Bar{D}(u, \nu)$ and $Q(u, \nu, p_{r}, p_{\varphi})$ are fixed by writing a 
particular ansantz for the potentials within the Hamiltonian expression for 
\eqn{h_eff_generic}, then matching this Hamiltonian through some gauge invariant quantity 
or through a canonical transformation to a PN-informed Hamiltonian. 

Traditionally, EOB  waveforms developed for observations 
in the LVK Collaboration, use in \eqn{h_eff_generic} a polynomial structure of 
the potentials almost entirely informed by PN-information.
For example, the $A(u, \nu)$ potential has the simple form in the nonspinning case,
\begin{multline}
	A(u, \nu) = 
	1 - 2 u + A_{2}(\nu)u^{2} + \\
	A_{3}(\nu)u^{3} + A_{4}(\nu)u^{4} + {\cal O}(u^{5} \ln u).
	\label{eq:a_potential_pn_form}
\end{multline}
In fact, this form becomes even simpler than \eqn{a_potential_pn_form} intially suggests,  
since the 1PN coefficient, $A_{2}(\nu)$, vanishes, whilst the coefficients $A_{3}(\nu)$ 
and $A_{4}(\nu)$ (i.e. up to 3PN) remain linear in $\nu$.
The potentials, $A(u, \nu)$, $\Bar{D}(u, \nu)$ and 
$Q(u, \nu, p_{r}, p_{\varphi})$, are known to
5.5PN order \cite{Bini:2020wpo}, but are missing two quadratic-in-$\nu$ coefficients in 
$A(u, \nu)$ and $\Bar{D}(u, \nu)$ at 5PN.

In general, the potential $Q(u, \nu, p_{r}, p_{\varphi})$ allows one to augment the EOB 
Hamiltonian through 3PN order, ensuring the original mapping 
outlined in \eqn{energy_map} remains unmodified. 
Furthermore, the mass-reduced potential
$\hat{Q}(u, \nu, \hat{p}_{r}, \hat{p}_{\varphi})$,
variably incorporates the mass-reduced radial and angular momenta, \emph{a priori}. 
Henceforth, we shall denote the $\hat{Q}$ function that only depends on the radial momentum 
as $\hat{Q}^{\rm DJS}(u,\nu,\hat{p}_{r})$, after the initials of the three authors of Ref.~\cite{Damour:2000we} 
who first used it.
It should be noted, however, that this freedom need not be fixed in this way and other choices
of this $\hat{Q}$ potential can be made (see, e.g.,  Refs.~\cite{Buonanno:2007pf, Damour:2002vi}).
Fixing this freedom means the reduced angular momentum only appears as 
$\sim \hat{p}_{\varphi}^{2}u^{2}$ and not explicitly within any of the potentials 
$A(u, \nu)$, $\Bar{D}(u, \nu)$ and $\hat{Q}^{\rm DJS}(u, \nu, \hat{p}_{r})$.

It was also shown in Ref.~\cite{Damour:2000we} that $\hat{Q}^{\rm DJS}(u, \nu, \hat{p}_{r})$ has the 
form:
\beq
	\hat{Q}^{\rm DJS}(u, \nu, \hat{p}_{r}) = 
	\hat{p}^{4}_{r} q_{4}(u) + \hat{p}^{6}_{r} q_{6}(u) + {\cal O}(\hat{p}^{8}_{r}),
	\label{eq:q_potential_pn_form}
\eeq
such that this mass-shell deformation term vanishes quartically\footnote{As in 
Ref.~\cite{Damour:2000we}, this can be seen by generalising the pseudo EOB 
``Hamilton-Jacobi" equation and comparing it to \eqn{mass_shell_condition}: 
$\mu^{2} + g^{\mu\nu}_{\rm eff}p_{\mu}p_{\nu} + 
A^{\mu\nu\gamma\delta} p_{\mu}p_{\nu}p_{\gamma}p_{\delta} + \dots = 0$.}
in the limit of $\hat{p}_{r} \rightarrow 0$.  Therefore, within the DJS gauge, one is forced to
incorporate $p^{2}_{\varphi}$-term into the geodesic part of the Hamiltonian.

In the DJS-gauge, the EOB-potentials can be rexpressed as a GSF-inspired expansion 
in the mass-ratio, $\nu$, at fixed $u$, whilst the form of the expansions appearing in 
Eqs.~(\ref{eq:a_potential_pn_form}) and (\ref{eq:q_potential_pn_form}) correspond to that of PN, (i.e., 
an expansion in $u$ for fixed $\nu$).  
If we again consider the potentials $A(u, \nu), Q^{\rm DJS}(u, \nu, p_{r}), \Bar{D}(u, \nu)$, 
we have linear-in-$\nu$ expansions about the Schwarzschild (nonspinning) limit:
\begin{subequations}
\begin{align}
	A(u, \nu) &= 1 - 2u + \nu a(u) + {\cal O}(\nu^{2}), \label{eq:a_potential_gsf_form}\\
	\Bar{D}(u, \nu) &= 1 + \nu \bar{d}(u) + {\cal O}(\nu^{2}), \label{eq:d_potential_gsf_form}\\
	Q^{\rm DJS}(u, \nu, {p}_{r}) &= \nu q(u) p^{4}_{r} + {\cal O}(\nu^{2}). 
	\label{eq:q_potential_gsf_form}
\end{align}
\end{subequations}
In particular if we focus on the first of these expanded potentials, \eqn{a_potential_gsf_form}, it has been
shown the function $a(u)$ can be directly related to the linear-in-$\nu$ correction to the gauge-invariant
quantity known as the \emph{Detweiler redshift} 
\cite{Detweiler:2008ft, Barack:2010tm, Barausse:2011dq, LeTiec:2011dp, Akcay:2012ea, Bini:2014zxa, Dolan:2014pja, Kavanagh:2015lva}.

The Detweiler redshift can be understood most simply from a SF perspective.  In this description,
the comparatively smaller secondary is considered to be a perturbation of the background (Schwarzschild) geometry 
such that the secondary moves on a geodesic of the effective (GSF) metric 
$\tilde{g}_{\mu\nu} = g_{\mu\nu} + h^{R}_{\mu\nu}$.  
The eponymous redshift variable is defined as
\beq
	z := (\tilde{u}^{t})^{-1} \diff{\tilde{\tau}}{t},
	\label{eq:detwieler_redshift}
\eeq
where $\tilde{u}^{\alpha}$ is the four-velocity normalised in the effective metric, $\tilde{g}_{\mu\nu}$, 
to linear-order in $\nu$: $\tilde{g}_{\mu\nu} \tilde{u}^{\mu} \tilde{u}^{\nu} = - 1 + {\cal O}(\nu)$.
Here $t$ and $\tilde{\tau}$ are the coordinate time and proper time in the effective metric respectively.
It is important to stress that since the Detweiler redshift is calculated in the effective metric,
this quantity cannot be interpreted as the physical gravitational redshift.
A more physical interpretation is that $z$ is proportional to the surface gravity\footnote{This 
relationship can be shown to be exact up to linear-order 
in the mass-ratio through the approach of matched asymptotic expansions 
\cite{PoundHorizon} and has been confirmed numerically through NR simulations 
\cite{Zimmerman:2016ajr}.} of the smaller BH's horizon 
\cite{PoundHorizon, Zimmerman:2016ajr}.

One can decompose $z$ to extract the linear-in-$\nu$ correction to the redshift,
\beq
	z(x) = \sqrt{1 - 3x} + \nu \Delta z(x) + {\cal O}(\nu^{2}),
	\label{eq:redshift_expansion_nu}
\eeq
where $\Delta z(x)$ (at fixed $x$) can be expressed as
\beq
	\Delta z(x) = \frac{x}{\sqrt{1 - 3x}} -\frac{1}{2}\sqrt{1 - 3x}\, h^{R}_{uu}(x).
	\label{eq:linear_in_nu_redshift}
\eeq
Here we have introduced the notation $h^{R}_{uu} := h^{R}_{\mu\nu} u^{\mu}u^{\nu}$, 
which is simply the double contraction of the regular metric perturbation with the 
four-velocity.  Furthermore $x$ is the gauge-independent frequency parameter, 
defined as $x := (M\Omega)^{2/3}$ where $\Omega$ is the orbital frequency.
It is important to note here that in SF calculations, the expansion of the redshift is viewed as 
an expansion in the (small) mass-ratio $q = m_{2}/m_{1}$ at fixed dimensionless frequency parameter 
$y = (m_{1}\Omega)^{2/3}$.  The equivalent expansion in the context of SF 
to \eqn{redshift_expansion_nu} is
\beq
	z(y) = \sqrt{1 - 3y} + q \Delta z^{\rm 1SF} (y) + {\cal O}(q^{2}).
	\label{eq:redshift_expansion_q}
\eeq
Although similar, the different form of the expansion in \eqn{redshift_expansion_nu},
as powers of $\nu$ instead of $q$, whilst holding $x$ fixed instead of $y$, means that the expansion 
coefficients differ even at linear-order in $q$.  Since $y$ and $x$ are related through $q$ and $\nu$, one finds
\beq
	\Delta z (x) = \Delta z^{\rm 1SF} (x) + \frac{x}{\sqrt{1 - 3x}},
\eeq
through ${\cal O}(q)$.

Since this expression does not involve any derivatives of the (regular) metric 
perturbation and, as first noted in \cite{Detweiler:2008ft}, is somewhat a 
gauge-invariant quantity, then this makes the redshift extremely appealing for use in 
synergistic studies with EOB.
Another feature of the Detweiler redshift is that one can relate $\Delta z$ to
the linear-in-$\nu$ correction to the $A(u, \nu)$ potential through the 
\emph{first law of binary--black-hole mechanics} \cite{LeTiec:2011ab,LeTiec:2011dp} such that
\beq
	a(x) = \Delta z(x)\sqrt{1 - 3x} - x \left( 1 + \frac{1 - 4x}{\sqrt{1 - 3	x}} \right).
	\label{eq:a_potential_redshift}
\eeq
One should note that this relation is expressed in terms of the gauge-independent $x$, to align this with the
rest of the section, rather the usual EOB inverse radius, $u$, which is gauge-dependent.  However, the two
expressions are trivially related since up to linear-order in $\nu$: $x = u + {\cal O}(\nu)$.

\subsection{Divergence at the light ring in DJS gauge}
As first observed in Ref.~\cite{Akcay:2012ea}, attempts to use the Detweiler redshift to inform the 
EOB potential, $a(x)$, encounter divergences towards the light-ring (LR), as $x \rightarrow 1/3$.
This is obvious from the last term in \eqn{a_potential_redshift} when we approach the LR, i.e. as $x \rightarrow 1/3$.
This divergence, however, is not limited to the form of the expression for $a(x)$.
As was first observed in Ref.~\cite{Akcay:2012ea} and in subsequent numerical SF codes, the 
contracted metric perturbation does indeed diverge at the light-ring.

Firstly, let's consider the divergence of the SF quantity, $h^{R}_{uu}$.  
This was shown to diverge as $h^{R}_{uu} \sim E_{\rm S}(x)^{3}$ as $x \rightarrow 1/3$, where $E_{\rm S}(x)$ 
is the specific energy of a particle in Schwarzschild spacetime:
\beq
	E_{\rm S}(x) = \frac{1 - 2x}{\sqrt{1 - 3x}}.
	\label{eq:1sf_specific_energy}
\eeq
As shown in Ref.~\cite{Akcay:2012ea} (and further supported in Ref.~\cite{Bini:2016cje}), the origin of this divergence 
can be traced to a simple heuristic argument: the metric perturbation is sourced not only by the rest mass $m_{2}$ of
the secondary, but by its conserved energy, $m_{2} E_{\rm S}(x)$.  Accordingly, $h^{R}_{\mu\nu} \sim E_{\rm S}(x)$
near the LR and the contraction with the particle's four-velocity, whose $u^{0}$-component also scales as $E_{\rm S}(x)$,
leads to the cubic divergence observed in $h^{R}_{uu}$.

The second divergence appearing in \eqn{linear_in_nu_redshift} is manifest as the $(1 - 3x)^{-1/2}$ factor
that appears in the bracketed term.  At first glance, this behaviour might suggest the presence of a physical
singularity in the EOB formalism.  However, as shown in Ref.~\cite{Akcay:2012ea}, the divergence is analogous to the coordinate
singularity in Schwarzschild spacetime and does not reflect any physical pathology.  More precisely, the singular
behaviour of the potential, $a(u)$, near the light-ring is a consequence of the choice of standard phase space coordinates
used in the DJS gauge. Ref.~\cite{Akcay:2012ea} showed that this divergence can be removed entirely by performing a 
suitable simpletic transformation, which amounts to a change of phase-space variables.  
Therefore, to develop an EOB Hamiltonian informed by SF, it appears essential to entirely forsake the 
conventional DJS gauge in favor of a gauge more suitable for the resummation of SF data.

\subsection{Post-Schwarzschild gauge}

In light of the divergences appearing in the DJS gauge, an alternative prescription initially suggested in 
Ref.~\cite{Akcay:2012ea} and further detailed in Ref.~\cite{Damour:2017zjx} proposes relaxing the gauge 
conditions of the DJS gauge.  
This modification allows the $\hat{Q}$-potential to scale as $\hat{Q} \sim \hat{p}_{\varphi}^{3}$ in the limit 
$\hat{p}_{\varphi} \rightarrow \infty$ and $\hat{p}_{r} \rightarrow 0$.  
In this modifcation to the usual gauge fixing, one fixes the radial metric coefficient in 
\eqn{eff_metric_line_element} to be equal to the Schwarzschild metric.
The corrections to the test-mass limit (TML) of the effective Hamiltonian are then incoporated into the 
$\hat{Q}$-potential such that
\beq
	\hat{H}^{\rm PS}_{\rm eff} = \sqrt{\hat{H}_{\rm S}^{2} + (1 - 2u)\hat{Q}^{\rm PS}(u, \hat{H}_{\rm S}, \nu)}.
	\label{eq:post_schwarzschild_hamiltonian}
\eeq
As in \cite{Antonelli:2019fmq, Antonelli:2019ytb}, we shall refer to this alternative gauge as a 
\emph{post-Schwarzschild (PS) gauge}.

In such a family of Hamiltonians, the information encapsulating the dynamics of the binary inspiral are now 
included inside the $\hat{Q}^{\rm PS}$-potential, rather than the usual $\hat{A}^{\rm DJS}$-potential.
Moreover, the $\hat{Q}^{\rm PS}$-potential in the PS Hamiltonian has a generic functional 
dependence on $\hat{H}_{\rm S}$.
One can therefore capture the divergent behaviour towards the LR through a leading-order cubic term in 
$\hat{H}_{\rm S}$, as was first indicated in Ref.~\cite{Damour:2017zjx}.

The $\hat{Q}^{\rm PS}$-potential can also be expressed analytically through 3PM+4PN order as
\begin{multline}
	\hat{Q}^{\rm PS}_{\rm 3PM+4PN} = u^{2}q_{\rm 2PM}(\hat{H}_{\rm S}, \nu) + u^{3}q_{\rm 3PM}(\hat{H}_{\rm S}, \nu) \\
	+ \Delta_{\rm 3PN}(u, \hat{H}_{\rm S}, \nu) + \Delta_{\rm 4PN}(u, \hat{H}_{\rm S}, \nu) + {\cal O}({\rm 5PN}),
	\label{eq:3PM+4PN_q_potential}
\end{multline}
where we include up to 4PN correction terms.  
The coefficients, $q_{\rm nPM}$, up to 3PM order are determined from calculation of the scattering angle 
\cite{Damour:2017zjx, Antonelli:2019ytb}.  
The higher-order PN correction terms, expressed as expansions in the PN parameters $u$ and $\hat{H}^{2}_{\rm S} - 1$,
are added as complimentary information to the purely PM expansions to ensure a match to the standard EOB DJS Hamiltonian 
up to 4PN order \cite{Damour:2015isa}.
To determine these terms, one takes the 4PN expansion of the 3PM Hamiltonian in \eqn{3PM+4PN_q_potential} and performs a canonical 
transformation. 
Then, the Hamiltonians are matched to the known 4PN EOB Hamiltonian in the DJS gauge, allowing one to solve for the unknown coefficients:
$\Delta_{\rm 3PN}(u, \hat{H}_{\rm S}, \nu)$ and $\Delta_{\rm 4PN}(u, \hat{H}_{\rm S}, \nu)$.
This information shall become important when considering augmenting the SF informed EOB
Hamiltonian beyond linear-order in the mass-ratio.

The concept of a redshift-informed potential, denoted as $\hat{Q}^{\rm PS}_{\rm SF}$\footnote{The potential, 
$\hat{Q}^{\rm PS}_{\rm SF}$, informed by $\Delta z$ data, was originally labeled  as 
$\hat{Q}^{\rm PS}_{\rm SMR}$, with ``${\rm SMR}$"  signifying ``small-mass-ratio". 
For greater clarity and to more accurately reflect its association with SF effects, 
we have adopted ``SF" in place of ``${\rm SMR}$", hence the notation $\hat{Q}^{\rm PS}_{\rm SF}$.}, and its 
subsequent implementation into a PS gauge Hamiltonian that seamlessly incorporates SF information without 
resulting in a divergence at the LR, was first introduced by Antonelli \emph{et al.} in Ref.~\cite{Antonelli:2019fmq}.
In this work, a PS gauge Hamiltonian was constructed from fitting regularised redshift data up from SF calculations up to the LR, but in such a manner that it remained analytic beyond it.
This Hamiltonian was then supplemented by incorporating additional PN information from the analytical expression 
of the $\hat{Q}^{\rm PS}$-potential in \eqn{3PM+4PN_q_potential}, thereby including beyond linear-order in the 
mass-ratio information as well as non-circular orbit terms.
In the following sections, we will briefly summarise its construction, whilst considering improvements on the 
original Hamiltonian.
A more complete deriviation, including additional corrections up to 3PN-order, can be found in the opening 
sections to Ref.~\cite{Antonelli:2019fmq}.

\section{Post-Schwarzschild gauge Hamiltonian with 4PN corrections}
\label{sec:ps_gauge_hamiltonian}
\subsection{Post-Schwarzschild gauge Hamiltonian informed from self-force}
\label{sec:ps_gauge_hamiltonian_self_force}
We shall consider the same ansatz for our $\hat{Q}^{\rm PS}_{\rm SF}$-potential as Ref.~\cite{Antonelli:2019fmq},
\begin{multline}
	\hat{Q}^{\rm PS}_{\rm SF}(u, \hat{H}_{\rm S}, \nu) = \nu \Big[ f_{0}(u) \hat{H}^{5}_{\rm S} \\
	+ f_{1}(u) \hat{H}^{2}_{\rm S} + f_{2}(u) \hat{H}^{3}_{\rm S} \ln \hat{H}^{-2}_{\rm S} \Big].
	\label{eq:q_potential_ansatz}
\end{multline}
The first term, proportional to $\hat{H}^{5}_{\rm S}$, is designed to capture the global 
divergence that appears in the redshift.  
As stated previously (see also footnote in Ref.~\cite{Antonelli:2019fmq}), a leading term proportional to 
$\hat{H}^{3}_{\rm S}$ would be sufficient to modify the divergence of $\sim (1 - 3x)^{-3/2}$, but this leads to 
pathological eccentric-like behaviour towards the LR.
Therefore, we make the same minimal correction as Ref.~\cite{Antonelli:2019fmq} that remedies this behaviour, 
whilst still capturing the divergence.
The second and third terms, proportional to $\sim (1 - 3x)^{-1/2}$ and $\sim \ln(1 - 3x)^{1/2}$ respectively, 
incoporate the problematic subleading terms appearing in the redshift and would potentially spoil the behaviour 
the Hamiltonian towards the LR.

To fix the functions $f_{i}(u)$ ($i \in 1, 2, 3$), one follows the procedure first outlined in 
Ref.~\cite{Barausse:2011dq}, and subsequently followed in Ref.~\cite{Antonelli:2019fmq}, by matching the binding 
energy from the EOB Hamiltonian and from SF calculations to linear-order in $\nu$ at fixed frequency.  
To compute the circular orbit binding energy from the PS-gauge Hamiltonian, one first determines the circular-orbit azimuthal momentum, $\hat{p}^{\rm circ}_{\varphi}$, by solving the Hamilton equation, 
\beq
	\dot{p}_{r} = - \pdiff{H_{\rm EOB}}{r}(r, p_{r} = 0, p_{\varphi}, \nu) = 0.
	\label{eq:pr_dot}
\eeq
This expression for $\hat{p}^{\rm circ}_{\varphi}$ is then used to express the gauge-dependent inverse radius $u$ 
and its gauge-dependent counterpart, $x$, to linear-order in $\nu$ through the second Hamilton equation for the 
circular orbit frequency:
\beq
	\Omega = \pdiff{H_{\rm EOB}}{p_{\varphi}}(r, p_{r} = 0, p_{\varphi}, \nu).
	\label{eq:Omega}
\eeq
The intermediate expressions for $\hat{p}^{\rm circ}_{\varphi}$ and $u^{\rm circ}(x, \nu)$ were given in 
Ref.~\cite{Antonelli:2019fmq} and are rewritten for completeness in Appendix~\ref{sec:sols_hamiltons_eqs}.
Together, the expressions for $\hat{p}^{\rm circ}_{\varphi}$ and $u^{\rm circ}(x, \nu)$ allow one to compute a 
gauge-invariant expression for the circular-orbit Hamiltonian, $\hat{H}_{\rm EOB}(x, \nu)$, and therefore the 
binding energy, $\hat{E}^{\rm EOB}_{\rm bind}(x, \nu)$ via
the definition:
\beq
	\hat{E}^{\rm EOB}_{\rm bind} := \frac{H_{\rm EOB} - M}{\mu}.
	\label{eq:binding_energy_eob}
\eeq
This yields an expression for the binding energy entirely in terms of $x, \nu$, and the functions 
$f_{i}(x)$ and their derivatives to linear-order in $\nu$.

In the same circular-orbit limit, one can express the binding energy in terms of the linear-in-$\nu$ correction 
to the redshift, $\Delta z(x)$ (see \eqn{redshift_expansion_nu}), through the first law of binary 
BH mechanics \cite{LeTiec:2011ab}.  
Explicitly, the binding energy is:
\begin{align}
	\hat{E}^{\rm SF}_{\rm bind}(x, \nu) &= \frac{1 - 2x}{\sqrt{1 - 3x}} - 1 + \nu \bigg[ \frac{\Delta z(x)}{2} + 
	\frac{(7 - 24x)x}{6(1 - 3x)^{3/2}} \nn \\
	&- \frac{x}{3}\Delta z^{\prime}(x) + \sqrt{1 - 3x} - 1 \bigg] + {\cal O}(\nu^{2}),
	\label{eq:binding_energy_sf}
\end{align}
where ${}^{\prime}$ denotes differentiation with respect to $x$.
Once one has the two binding energies, it is possible to link the redshift correction appearing in 
\eqn{binding_energy_sf} to the $\hat{Q}^{\rm PS}_{\rm SF}$-potential through the functions $f_{i}(u)$.
In order to do this within an EOB model, one needs to produce a suitable analytical fit for the correction to 
the redshift, $\Delta z$.
Furthermore, as noted previously, one needs to be careful of the actual divergent behaviour of the 
SF results for the Detweiler redshift.
Thus, Ref.~\cite{Antonelli:2019fmq} decomposes $\Delta z(x)$ further to anticipate the form of the fit to high-precision--numerical SF results such that
\beq
	\Delta z(x) = \frac{\Delta z^{(0)}(x)}{1 - 3x} + \frac{\Delta z^{(1)}(x)}{\sqrt{1 - 3x}} + \frac{\Delta 
	z^{(2)}(x)}{1 - 3x}\ln E^{-2}_{\rm S}(x).
	\label{eq:redshift_analytical_fit}
\eeq
The functions, $\Delta z^{(i)}(x)$, ($i \in 0, 1, 2$) are the actual pieces fitted to the numerical redshift 
data, whilst at the same time remaining analytic so as to preserve the analytic nature of the EOB model beyond 
the LR.  To ensure these functions enter the potential in such a way to preseve the analyticity of the 
PS-gauge Hamiltonian, we separate the functions $f_{i}(x)$ with the ansatz,
\beq
	f_{i}(x) = \tilde{f}_{i}(x) + \sum^{j = 2}_{j = 0} f^{(j)}_{i}(x)\Delta z^{(j)}(x), \quad i \in 0,1,2.
	\label{eq:f_ansatz}
\eeq
By imposing smoothness at the LR, by which we mean they should be free of spurious terms like $(1 - 3x)^{-1/2}$ 
or $\ln E^{-2}_{\rm S}$, we obtain solutions for the coefficients $\tilde{f}_{i}(x)$ and $f^{(j)}_{i}(x)$, with 
the rest of the function determined by the form of the fit to the redshift.
The non-zero solutions, given first in Ref.~\cite{Antonelli:2019fmq}, read
\begin{subequations}
	\begin{align}
	\tilde{f}_0(x) &= -\frac{x(1-3x)\left(1-4x\right)}{(1-2x)^5}\,\label{eq:f0_tilde},\\
	\tilde{f}_1(x) &= -\frac{x}{(1-2 x)^2}\,\label{eq:f1_tilde},\\
	f_0^{(0)}(x) &= \frac{1-3x}{(1-2x)^5}\,,\\
	f_1^{(1)}(x) &= \frac{1}{(1-2x)^2}\,,\\
	f_2^{(2)}(x) &= \frac{1}{(1-2x)^3}\,.
	\end{align}
\end{subequations}

Let us now return to the remaining part of the functions entering the PS-gauge Hamiltonian, the redshift fitting functions $\Delta z^{(i)}(x)$.  
There are two main considerations one must consider in the numerical fitting of these functions:  
\begin{enumerate*}[label=(\roman*)] \item encoding the precise numerical information from
redshift data without including the divergence at the LR, and
\item the lack of numerical SF data beyond the LR.
\end{enumerate*}
With this in mind, we follow the fitting procedure first outlined in the Appendix of Ref.~\cite{Antonelli:2019fmq} to 
obtain a robust and appropiate fit by utilising a 21.5 PN expansion of $\Delta z$, first obtained in 
Kavanagh \emph{et al.}~in Ref.~\cite{Kavanagh:2015lva}.
The analytical nature of this expression allows one to minimize the need for a large number of fitting parameters,  
as well as allowing the numerical data to be mostly responsible for augmenting the fit in the strong-field regime.

\begin{figure}[h]
	\includegraphics[width=\linewidth]{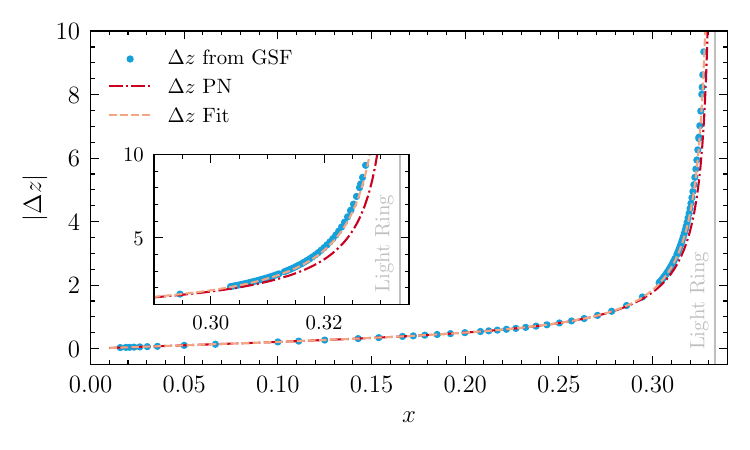}
	\caption{We demonstrate the semi-analytical fitting of the Detweiler redshift from numerical
	GSF data towards the LR as a function of the gauge-invariant radius, $x$.  The raw numerical GSF
	data for the linear-in-$\nu$ correction to the redshift is shown by the blue circles up to 
	$(1 - 3x) \simeq 4 \times 10^{-5}$.  Each data point has been calculated using mode-sum regularization 
	with 120 $\ell$-modes.  The red dash-dotted line shows the fit using only the 21.5PN expression from
	Ref.~\cite{Kavanagh:2015lva}, $Z_{\rm PN}(x)$, i.e. \eqn{redshift_fit_form} with $Z_{\rm fit}(x)$ set
	to zero.  The orange dashed line, on the other hand, indicates the correction to
	the redshift as calculated using our numerical fit in conjunction with $Z_{\rm PN}(x)$ (i.e.,
	\eqn{redshift_fit_form} including \eqn{redshift_fitting_function}).
	One can see, especially from the inset, that towards the LR the accuracy of our fit to GSF data
	is greatly improved by the inclusion of the fit, $Z_{\rm fit}(x)$.}
	\label{fig:redshift_fit}
\end{figure}

The fit takes the same form as outlined in Ref.~\cite{Antonelli:2019fmq}, but we find it clearer to be more explicit by writing it in the following manner:
\beq
	\Delta z = Z_{0}(x) + 2x^{3}\frac{(1 - 2x)^{3}}{1- 3x}Z_{\rm PN}(x)\left[1 + \alpha(x)Z_{\rm fit}(x)\right],
	\label{eq:redshift_fit_form}
\eeq
where $Z_{0}(x)$ is the leading term that is fixed such that the leading terms $\tilde{f}_0(x)$ and $\tilde{f}_1(x)$ in Eqs.~(\ref{eq:f0_tilde}) and (\ref{eq:f1_tilde}) are exactly cancelled and do not enter into the final Hamiltonian.
The additional function, $\alpha(x)$, that multiples $Z_{\rm fit}$ in \eqn{redshift_fit_form}, is an attenuation function designed to ensure that in the weak field the entire fitting is dominated by the form of the high-order PN resummation, $Z_{\rm PN}(x)$, which we now discuss shortly.
It is preferable, since this function inevitably enters into the expression for the Hamiltonian, that it is smooth and
differentiable throughout the entire domain, and hence this is realised by taking this form to be an exponential such that
\beq
	\alpha(x) = \exp \left[ \frac{4 - x^{-2}}{6} \right].
	\label{eq:attenuation_function}
\eeq
We define $Z_{\rm PN}(x)$ by starting from the 21.5PN expression for $\Delta z$, as given in Ref.~\cite{Kavanagh:2015lva},
setting $Z_{\rm fit}(x) = 0$ in order to analytically solve for $Z_{\rm PN}(x)$. Once inverted, the solution is re-expanded
in terms of $x$ (and $\log x$) to the same PN-order as the starting series. To ensure the Hamiltonian remains finite
at the horizon (i.e., $x = 1/2$), we pull out the divergent factor $(1 - 2x)^{2}$ and resum the remaining series
so that $Z_{\rm PN}(x) \rightarrow 0$ as $x \rightarrow 1/2$.
Omitting the full expression for the sake of brevity, the final resummed form reads
\beq
	Z_{\rm PN}(x) = (1-2x)^{2}\sum_{i, j}b_{(i, j)}x^{i/2}\log^{j} x,
	\label{eq:redshift_pn_form}
\eeq
where the first few non-zero coefficients are given by
\begin{subequations}
\begin{align}
	b_{(0, 0)} &= 1, \\
	b_{(2, 0)} &= \frac{1}{192}(4640 - 123\pi^{2}), \\
	b_{(4, 0)} &= \frac{2126}{15} + \frac{64\gamma_{\rm E}}{5} - \frac{3301\pi^{2}}{1024} + \frac{128\log 2}{5} \\
	b_{(4, 1)} &= \frac{32}{5}.
	\label{eq:redshift_pn_coefficients}
\end{align}
\end{subequations}
Here, $\gamma_{\rm E} \simeq 0.5772$, is the Euler's constant.
The final piece of the fitting function, namely $Z_{\rm fit}(x)$, is a combination of five polynomial and logarithmic terms given by,
\beq
	Z_{\rm fit}(x) = c_{0} + c_{1}\beta + c_{2}\beta^{4} + (c_{3}\beta + c_{4}\beta^{4})\log \left[ \frac{1 - 3x}{(1 - 2x)^2} \right],
	\label{eq:redshift_fitting_function}
\eeq
with the polynomial function $\beta := 9x(1 - 3x)(1 - 2x)$, and $c_{i}$ ($i \in [0, 4]$) being abitrary coefficients to be determined by
linearly fitting to numerical data.
To determine the fitting coefficients $c_{i}$, Ref.~\cite{Antonelli:2019fmq} used the high-precision SF  
code developed by one of the authors in Ref.~\cite{vandeMeent:2015lxa}.  
This code can be used to construct $h^{R}_{uu}$ using the method of mode-sum regularisation with the use of the 
arbitrary precision arithmetic to enable the production of high-precision results. 
However, as noted in Ref.~\cite{Akcay:2012ea}, near the LR the precision is ultimately limited by the number 
of $\ell$-modes that are included. 
Due to relativistic beaming effects one needs to include order $\sim (1-3x)^{-1}$ $\ell$-modes to achieve 
convergence of the mode sum for gauge invariant quantities such as the redshift the LR.
Here, we use the same data and fitting as in \cite{Akcay:2012ea, Antonelli:2019fmq}, with numerical data obtained 
for $\Delta z$ up to $(1 - 3x) \simeq 4 \times 10^{-5}$ with $120\ \ell$-modes.  
The optimal coefficients, $c_{i}$, are found to be 
\begin{subequations}
\begin{align}
	c_{0} &=  0.5559473282711097 \\
	c_{1} &=  -2.5898685656267251 \\
	c_{2} &=  31.1449861760975693 \\
	c_{3} &=  2.4401152757341507 \\
	c_{4} &= -179.1758186033215239
	\label{eq:redshift_fitting_coefficients}
\end{align}
\end{subequations}
One can then combine all of these results together to obtain the final expressions, $f_{i}(x)$, that enter into the $\hat{Q}^{\rm PS}_{\rm SF}$ potential, in \eqn{q_potential_ansatz}.
The coefficient of $H_{\rm S}^{2}$, $f_{1}(x)$, actually vanishes whilst the rest the two remaining functions can be written in the succinct form:
\begin{subequations}
\begin{align}
	f_{0}(x) &= \frac{1-3x}{(1-2x)^{2}} Z_{\rm PN}(x) \left[ 1 + \alpha(x)(c_{0} + c_{1} \beta + c_{2} \beta^{4}) \right],\\
	f_{2}(x) &= (1-2x)^{2} Z_{\rm PN}(x)\left[ 1 + \alpha(x)(c_{3} \beta + c_{4} \beta^4) \right].
\end{align}
\end{subequations}
The final semi-analytical fit is demonstrated for the linear-in-$\nu$ correction to the redshift in
\fig{redshift_fit}.

This completes the SF-informed piece of the potential, $\hat{Q}^{\rm PS}_{\rm SF}$, but Ref.~\cite{Antonelli:2019fmq} showed that
information beyond the circular-orbit, linear-in-mass-ratio limit can be included to enhance the $\hat{Q}^{\rm PS}$ based on previous
studies up to 4PN-order \cite{Antonelli:2019ytb}, see \eqn{3PM+4PN_q_potential}.

\subsection{3PN and 4PN corrections to the post-Schwarzschild gauge Hamiltonian}
The main difficulty is that arbitrarily adding PN terms to the $\hat{Q}^{\rm PS}_{\rm SF}$-potential introduces double counting 
of the PN information already included implicitly into our Hamiltonian.
In particular, since in the PS-gauge the momenta enters through the factors of 
$\hat{H}_{\rm S} = \hat{H}_{\rm S}(\hat{u}, \hat{p}_{r}, \hat{p}_{\varphi})$, then one cannot split the potential into pieces that are 
purely circular and non-circular contributions.
Thus, any addition of PN-terms that are to linear-order-in-$\nu$ will affect the precise matching done for the binding energies for 
$\hat{Q}^{\rm PS}_{\rm SF}$ in the circular orbit limit.
To augment the Hamiltonian further one splits the potential into a linear-order-in-$\nu$ piece, $\hat{Q}^{\rm PS}_{\rm SF}$, and then 
an additional term, $\Delta\hat{Q}^{\rm PS}$, informed by additional PN-terms:
\beq
	\hat{Q}^{\rm PS}_{{\rm SF}-n{\rm PN}} = \hat{Q}^{\rm PS}_{\rm SF} + \Delta\hat{Q}^{\rm PS}_{n{\rm PN}}.
	\label{eq:q_potential_sf_npn}
\eeq

In the same manner as Ref.~\cite{Antonelli:2019fmq}, this term has to be broken down further such that
\beq
	\Delta\hat{Q}^{\rm PS}_{n{\rm PN}} := \Delta\hat{Q}^{\rm PS}_{n{\rm PN (extra)}} + \Delta\hat{Q}^{\rm PS}_{n{\rm PN (count)}},
	\label{eq:q_potential_pn_term}
\eeq
where $\Delta\hat{Q}^{\rm PS}_{n{\rm PN (extra)}}$ captures the additional PN-terms introduced up to $n$PN-order, whilst 
$\Delta\hat{Q}^{\rm PS}_{n{\rm PN (count)}}$ is a counter-term designed specifically to cancel the linear-in-$\nu$ PN terms 
that are introduced in the first term that would spoil the agreement of the binding energy to linear-order in $\nu$.

To calculate $\Delta\hat{Q}^{\rm PS}_{n{\rm PN (extra)}}$, we find the difference between the PN-expansion of 
$\hat{Q}^{\rm PS}_{\rm 3PM+4PN}$ in \eqn{3PM+4PN_q_potential} with coefficients taken from 
Ref.~\cite{Antonelli:2019ytb} and 
$\hat{Q}^{\rm PS}_{\rm SF}$.
Given that $\hat{Q}^{\rm PS}_{\rm SF}$ is semi-analytical, incorporating SF-information through fitting to 
numerical data rather than solely using analytical PN expressions for $\hat{Q}^{\rm PS}$, we calculate a 3PN 
expansion of $\hat{Q}^{\rm PS}_{\rm SF}$. 
This is done by replacing the numerical fitting for $\Delta z$ with PN expansions of the redshift functions $
\Delta z^{(i)}$, which are derived from by matching the 3PN expansion of Detweiler's redshift invariant obtained 
in Ref.~\cite{Kavanagh:2016idg} with 
\eqn{redshift_analytical_fit}.
To obtain such expansions, one expands in the PN parameters $u$ and 
$Y := \hat{H}^{2}_{\rm S} - 1 \sim {\cal O}(1/c^{2})$.
In Ref.~\cite{Antonelli:2019fmq}, the mixed SF-PN potential is considered up-to-3PN order, but we proceed to 
include the same 3PN terms \emph{and} 4PN corrections.
It is important to note there is an important distinction that $\Delta\hat{Q}^{\rm PS}_{3{\rm PN (extra)}}$ is 
valid for generic orbits,
whilst at one PN-order higher, $\Delta\hat{Q}^{\rm PS}_{4{\rm PN (extra)}}$ is only valid for near-circular 
orbits.
The 4PN tail-term corrections to the potential are only valid in the near-circular-orbit-regime 
since the corrections are computed through an 
expansion in $\hat{p}_{r}$ up to ${\cal O}(\hat{p}^{6}_{r})$, which can be thought of as an small eccentricity 
expansion.
In this work, we utilize the 3PN and 4PN potentials, $\hat{Q}^{\rm PS}_{\rm 3PM+3PN}$ and $\hat{Q}^{\rm PS}_{\rm 
3PM+4PN}$ first 
presented in Ref.~\cite{Antonelli:2019ytb}, 
with the $\hat{Q}^{\rm PS}_{\rm 3PM+3PN}$ used in the model in Ref.~\cite{Antonelli:2019fmq}.
The additional PN-terms up to 4PN are given 
by\footnote{In Ref.~\cite{Antonelli:2019ytb}, there is a typo in Eq.~(A10).  The $\nu^{2}$-coefficient,
${123}/{54}\pi^{2}$, should read ${123}/{64}\pi^{2}$ as in our expression for \eqn{Q4PN_extra}.
We thank Mohammed Khalil for confirming this typographical error.}:
\begin{subequations}
\begin{align}
	&\Delta\hat{Q}^{\rm PS}_{3{\rm PN (extra)}} = 3\nu u^2 Y+\bigg(3\nu-\frac{9}{4} \nu^2 \bigg) u^2Y^2+3 \nu u^3 
	\nn\\
	&+\bigg(22\nu-\frac{23}{4} \nu^2 \bigg) u^3Y+\bigg(16 \nu-\frac{7}{2} \nu^2 \bigg)  u^4\,,
	\label{eq:Q3PN_extra}
\end{align}
\begin{widetext}
\begin{align}
	&\Delta\hat{Q}^{\rm PS}_{4{\rm PN (extra)}} = 
	3 u^3 \nu + 3 u^2 Y \nu + u^3 Y \bigg(22 \nu - \frac{23 \nu^2}{4}\bigg)
	+ u^4 \bigg(16 \nu - \frac{7 \nu^2}{2}\bigg) + u^2 Y^2 \bigg(3 \nu - \frac{9 \nu^2}{4}\bigg) \nn\\
	&+ u^2 Y^3 \bigg(-\frac{4135 \nu}{48} - \frac{27 \nu^2}{16} + \frac{15 \nu^3}{8} - \frac{147432}{5} \nu \log 
	2 + \frac{1399437}{160} \nu \log 3 + \frac{1953125}{288} \nu \log 5 \bigg) \nn\\
	&+ u^3 Y^2 \bigg(-\frac{304853 \nu}{1200} - \frac{445 \nu^2}{16} + \frac{49 \nu^3}{8}
	- \frac{14099512}{225} \nu \log 2 + \frac{14336271}{800} \nu \log 3
	+ \frac{4296875}{288} \nu \log 5\bigg) \nn\\
	&+ u^5 \bigg(-\frac{283739 \nu}{1800}
	+ \frac{296 \gamma_{\rm E}\, \nu}{15} - \frac{13921 \pi^2 \nu}{6144} \nn\\
	&- \frac{2387 \nu^2}{24} + \frac{205 \pi^2 \nu^2}{64} + \frac{9 \nu^3}{4}
	- \frac{256216}{25} \nu \log 2 + \frac{1061181}{400} \nu \log 3
	+ \frac{390625}{144} \nu \log 5 + \frac{148}{15} \nu \log u\bigg) \nn\\
	&+ u^4 Y \bigg(-\frac{135239 \nu}{450} + \frac{296 \gamma_{\rm E}\, \nu}{15} - \frac{13921 \pi^2 \nu}{6144} 
	\nn\\
	&- \frac{405 \nu^2}{4} + \frac{123 	\pi^2 \nu^2}{64} + \frac{13 \nu^3}{2} - \frac{9771016}{225} \nu \log 2 
	+ \frac{1182681}{100} \nu \log 3 + \frac{390625}{36} \nu \log 5 + \frac{148}{15} \nu \log u\bigg) \,.
	\label{eq:Q4PN_extra}
\end{align}
\end{widetext}
\end{subequations}

The second, correctional term, $\Delta\hat{Q}^{\rm PS}_{n{\rm PN (count)}}$, then ensures the matching done at the level of the
binding energies through the redshift invariant is not disrupted by the additional PN terms in 
$\Delta\hat{Q}^{\rm PS}_{n{\rm PN (extra)}}$.
To do this in a systematic way, this term is choosen to start at $(n + 1)$-PN order higher than the 
$\Delta\hat{Q}^{\rm PS}_{n{\rm PN (count)}}$ term.
Whereas the calculation of these particular terms is very much agnostic of the particular PN-order one is working at, the form of the 
counter term, however, is different depending on the particular PN-order.
The main difference is that at 4PN, the counter term must include logarithmic terms that only begin to enter 
$\Delta\hat{Q}^{\rm PS}_{n{\rm PN (extra)}}$ at 4PN-order.

At 3PN, the counter-term was first given in Ref.~\cite{Antonelli:2019fmq} by
\begin{align}
	\Delta\hat{Q}^{\rm PS}_{3{\rm PN (count)}} = \nu \left( q_{3{\rm PN}}^{(3,2)} u^{3}Y^{2} + q_{3{\rm PN}}^{(4,1)} u^{4}Y 
	+ q_{3{\rm PN}}^{(5,0)} Y^{5} \right),
	\label{eq:Q3PN_count}
\end{align}
where
\begin{align}
	q_{3{\rm PN}}^{(3,2)} = 9,\quad q_{3{\rm PN}}^{(4,1)} = 96,\quad q_{3{\rm PN}}^{(5,0)} = 112.
	\label{eq:Q3PN_count_coefficients}
\end{align}
Then, at 4PN, the counter-term now starts at 5PN and now includes logarithmic terms such that
\begin{multline}
	\Delta\hat{Q}^{\rm PS}_{4{\rm PN (count)}} = \nu \sum^{i = 3}_{i = 0} \bigg( q_{4{\rm PN}}^{(6 - i,\, i)} u^{(6 - i)} Y^{i} \\
	+ q_{4{\rm PN}, \log}^{(6 - i,\, i)} u^{(6 - i)} Y^{i} \log u \bigg),
	\label{eq:Q4PN_count}
\end{multline}
where the coefficients for the non-logarithmic terms are given by,
\begin{align}
	q_{4{\rm PN}}^{(0,6)} &= -\frac{82139}{450} + \frac{1184 \gamma_{\rm E}}{15} - \frac{13921 \pi^2}{1536} \nn \\
	&- \frac{1024864 \log 2}{25} + \frac{1061181 \log 3}{100} + \frac{390625 \log 5}{36}, \nn \\
	q_{4{\rm PN}}^{(1,5)} &= -\frac{92699}{120} + \frac{296 \gamma_{\rm E}}{5} - \frac{13921 \pi^2}{2048} \nn \\
	&- \frac{7355624 \log 2}{45} + \frac{3572343 \log 3}{80} + \frac{1953125 \log 5}{48}, \nn \\
	q_{4{\rm PN}}^{(2,4)} &= -\frac{492767}{600} - \frac{16310992 \log 2}{75} \nn \\
	&+ \frac{25002999 \log 3}{400} + \frac{7421875 \log 5}{144}, \nn \\
	q_{4{\rm PN}}^{(3,3)} &= -\frac{4135}{16} - \frac{442296 \log 2}{5} + \frac{4198311 \log 3}{160} \nn \\
	&+ \frac{1953125 \log 5}{96},
	\label{eq:Q4PN_count_coefficients}
\end{align}
and the coefficients for the logarithmic terms are given by,
\begin{align}
	q_{4{\rm PN}, \log}^{(0,6)} &= \frac{592}{15}, \quad
	q_{4{\rm PN}, \log}^{(1,5)} = \frac{148}{5}, \nn \\
	q_{4{\rm PN}, \log}^{(2,4)} &= q_{4{\rm PN}, \log}^{(3,3)} = 0.
\end{align}

\section{Inspiral-merger-ringdown model}
\label{sec:imr_model}

\subsection{Summary of \texttt{SEOBNRv5} gravitational modes}

In constructing a full, semi-analytical EOB model to produce waveforms for an inspiral-merger-ringdown (IMR) model, one requires three necessary components.
The first is a two-body Hamiltonian which encodes the conservative EOB dynamics.
It is from this Hamiltonian that one obtains the equations of motion (EOMs).
In this work, we completely codify this conservative sector with our PS-gauge Hamiltonian informed with SF data and then
additionally supplemented with PN-information that was presented in the previous section.
The other two components required to complete our model are the EOB \emph{radiation-reaction forces} and the \emph{gravitational modes}. 
To complete our model, we employ the \texttt{SEOBNRv5} waveform model built for the fourth LVK observing run (O4)~\footnote{\texttt{SEOBNRv5} can be 
accessed via the \texttt{pySEOBNR} Python package, which is hosted at \href{https://git.ligo.org/waveforms/
software/pyseobnr}{\texttt{git.ligo.org/waveforms/software/pyseobnr}}. For stable releases of \texttt{pySEOBNR}, users can download them from the Python Package 
Index (PyPI) and install using the command:~\texttt{pip install pyseobnr}.}~\cite{LVKO4}.
\texttt{SEOBNRv5} is an IMR multipolar waveform model developed for quasicircular, spinning, nonprecessing black hole binaries and was introduced in a series of
articles \cite{Khalil:2023kep, Pompili:2023tna, vandeMeent:2023ols, Mihaylov:2023bkc}.
\texttt{SEOBNRv5} is particularly advantageous for our use since the software package for its use, \texttt{pySEOBNR}, has been built in a very modular fashion
 in the \texttt{Python} language.  
A detailed overview of this software package can be found in Ref.~\cite{Mihaylov:2023bkc}.
The \texttt{pySEOBNR} framework is suitably flexible in that the radiation-reaction (RR) forces and gravitational modes are 
implemented independently of the specific form of the EOB Hamiltonian.  
Therefore this allows the \texttt{SEOBNRv5} components used in this part of the model to remain unchanged, 
whilst one is free to insert an entirely new Hamiltonian.

In the following section, we briefly review the relevant \texttt{SEOBNRv5} components. 
For a more complete and detailed discussion, we refer the reader to 
Refs.~\cite{Khalil:2023kep, Pompili:2023tna, vandeMeent:2023ols}.
In the \texttt{SEOBNRv5} model, the complete binary coalescence process, including the inspiral, plunge, merger, and ringdown phases, is described through the gravitational polarizations in the observer's frame:
\begin{align}
    h(t;\iota,\varphi_0) &= h_{+}(t;\iota,\varphi_0) - i h_{\times}(t;\iota,\varphi_0)\,, \nonumber \\
    &= \sum_{\ell=2}^\infty \sum_{m=-\ell}^\ell {}_{-2}Y_{\ell m}(\iota, \varphi_0) h_{\ell m}(t; \bm{\lambda}),
\end{align}
where $\iota$ is the inclination angle of the binary's orbit relative to the observer, and $\varphi_{0}$ is the azimuthal orientation. 
The functions ${}_{-2}Y_{\ell m}(\iota, \varphi_0)$ are the -2 spin-weighted spherical harmonics and $\bm{\lambda}$ signify the intrinsic parameters of the compact binary source.
In reality, since the waveform scales with the total mass, $M$, the waveform modes $h_{\ell m}(t; \bm{\lambda})$ are simply a function of the mass ratio $q$ and the spins: $\chi_{1}$ and $\chi_{2}$.
The waveform model simplifies for spin aligned/anti-aligned binaries with $h_{\ell m} = (-1)^\ell h_{\ell-m}^{*}$, reducing the number of modes necessary for analysis.

Within this framework, the seven dominant modes considered are $(\ell, m) = (2,2),\, (2,1),\, (3,3),\, (3,2),\, (4,4),\, (4,3),$ and $(5,5)$. The evolution of each mode through the coalescence stages is given by:
\begin{equation}
    h_{\ell m}(t) =
    \begin{cases}
        h_{\ell m}^{\rm{insp-plunge}}(t), & t < t_{\rm{match}}^{\ell m} \\
        h_{\ell m}^{\rm{merger-RD}}(t), &  t > t_{\rm{match}}^{\ell m}
    \end{cases},
\end{equation}
where the matching times $t_{\rm{match}}^{\ell m}$ are critical for transitioning between phases:
\begin{equation}
    t_{\rm{match}}^{\ell m} =
    	\begin{cases}
			t_{\rm{peak}}^{22}, &(\ell, m) = (2,2),(3,3),(2,1), \\
			& \qquad\quad\,\; (4,4),(3,2),(4,3) \\
			t_{\rm{peak}}^{22}-10 M,  &(\ell, m) =(5,5),
		\end{cases}
		\label{eq:matching_times}
\end{equation}
Here $t_{\rm{peak}}^{22}$ as the peak of the (2,2)-mode amplitude, and establishes a relationship with the remnant-BH innermost stable circular orbit (ISCO):
\begin{equation}
    t_{\rm{peak}}^{22} = t_{\rm{ISCO}} + \Delta t_{\rm{NR}},
    \label{eq:t_22_peak}
\end{equation}
where $t_{\rm{ISCO}}$ refers to the ISCO time for a Kerr black hole characterized by the remnant's final mass and spin \cite{Bardeen:1972fi}.
$\Delta t_{\rm{NR}}$ is a calibration parameter for aligning the EOB and NR (2,2)-mode peaks at merger.
We shall return to this calibration for our model in \secref{delta_t_22_calibration}.
From \eqn{matching_times}, one observes a different choice of matching time is made for the $(5,5)$-mode.  
This distinct choice is discussed in detail in Refs.~\cite{Cotesta:2018fcv, Pompili:2023tna}.

The gravitational waveforms for the inspiral and plunge are formulated as:
\begin{equation}
    h_{\ell m}^{\rm{insp-plunge}}(t) = h_{\ell m}^{\rm{F}}(t) N_{\ell m}(t),
    \label{eq:inspiral_plunge_form}
\end{equation}
where $h_{\ell m}^{\rm{F}}(t)$ aggregates the PN expanded gravitational wave modes for circular orbits in a factorized 
manner~\cite{Damour:2002vi, Damour:2008te, Buonanno:2009qa}, and $N_{\ell m}(t)$ applies non-quasicircular (NQC) corrections to account 
for radial effects during the plunge \cite{Damour:2007xr, Damour:2007yf, Damour:2008gu, Pan:2010hz}.

The complete factorized inspiral waveform, $h_{\ell m}^{\rm F}$, is described by the following product~\cite{Damour:2008gu}:
\beq
	h_{\ell m}^F = h_{\ell m}^{\rm N} \hat{S}_{\ell m} T_{\ell m} f_{\ell m} e^{i \delta_{\ell m}}.
\eeq
The leading term, $h_{\ell m}^{\rm N}$, represents the Newtonian contribution to the waveform, providing a foundation upon which further relativistic corrections are built. 
It is explicitly given by:
\beq
	h_{\ell m}^{\rm N} = \frac{\nu M}{d_L} n_{\ell m} c_{\ell+\epsilon_{\ell m}}(\mu)v_{\varphi}^{\ell+\epsilon_{\ell m}} 
	Y_{\ell-\epsilon_{\ell m},-m}\left(\frac{\pi}{2},\varphi\right),
\eeq
where $d_{L}$ the luminosity distance, and $Y_{\ell m}$ the spherical harmonics. 
The coefficients $n_{\ell m}$ and $c_{k}(\nu)$ are functions that modulate the amplitude based on the orbital dynamics and read:
\beq
		n_{\ell m} = \left\{\begin{aligned}
		& \frac{8\pi (i m)^\ell}{(2\ell+1)!!}\sqrt{\tfrac{(\ell+1)(\ell+2)}{\ell(\ell-1)}}, &&\text{$\ell+m$ is even},\\
		& \frac{-16i \pi (i m)^\ell}{(2\ell+1)!!}\sqrt{\tfrac{(2\ell+1)(\ell+2)(\ell^2-m^2)}{(2\ell-1)(\ell+1)\ell(\ell-1)}},
		&&\text{$\ell+m$ is odd},
	\end{aligned} \right.
\eeq
with
\beq
	c_{k}(\nu) = \left(\tfrac{1-\sqrt{1-4\nu}}{2} \right)^{k-1}+(-1)^k \left(\tfrac{1+\sqrt{1-4\nu}}{2} \right)^{k-1}.
\eeq
The parity $\epsilon_{\ell m}$ of the mode, is defined as
\beq
	\epsilon_{\ell m} = \begin{cases}
	0, &\ell+m \text{ is even},\\
	1, &\ell+m \text{ is odd}.
	\end{cases}
\eeq
Futhermore, the velocity $v_{\varphi}$ is derived from the orbital dynamics as specified by $H_{\rm EOB}$, with $\Omega$ and 
$r_{\Omega}$ denoting the angular and radial frequencies, respectively:
\beq
	\Omega = \frac{\partial H_{\rm EOB}}{\partial p_{\varphi}}, \quad v_{\varphi} = M\Omega r_\Omega, \quad \frac{1}{r_\Omega^{3/2}} 
	= \left.\frac{\partial H_{\rm EOB}}{\partial p_{\varphi}}\right|_{p_r=0}.
\eeq
The term $\hat{S}_{\ell m}$ acts as an effective source term for the waveform, with its form determined by the parity of the mode:
\beq
	\hat{S}_{\ell m} = \begin{cases}
	\hat{H}_{\rm eff}, & \ell+m \text{ is even},\\
	\sqrt{x} \hat{p}_{\varphi} & \ell+m \text{ is odd},
	\end{cases}
\eeq
where $v_{\Omega} = (M\Omega)^{1/3}$, and $\hat{H}{\rm eff}$ and $\hat{p}_{\varphi}$ denote the reduced effective energy and 
the orbital angular momentum, respectively, each normalised by its Newtonian counterpart.

The component $T_{\ell m}$ incorporates the leading-order tail effects through a resummation of an infinite number of leading logarithms \cite{Blanchet:1997jj}, 
such that
\beq
	T_{\ell m} = \frac{\Gamma(\ell+1-2i\hat{\Omega})}{\Gamma(\ell+1)}e^{\pi m \hat{\Omega}} \left(\frac{4m M\Omega}{\sqrt{e}}\right)^{2i m\hat{\Omega}},
\eeq
where $\hat{\Omega} = \Omega H_{\rm EOB}$ represents the normalized orbital frequency by the total energy, and $\Gamma(.)$ is the Euler gamma function. 

Finally, the amplitude and phase $ f_{\ell m} $ and $\delta_{\ell m} $ are polynomials in the velocity, whose coefficients are determined by requiring that the 
factorized modes agree with the expanded PN expressions.
For brevity, we do not reproduce the full expressions here and instead refer the reader to Appendix B 
of Ref.~\cite{Pompili:2023tna}, where they are given in full for $\texttt{SEOBNRv5}$.
In the nonspinning case, the amplitude is futher resummed according to $f_{\ell m} = (\rho_{\ell m})^{\ell}$,
in order to mitigate the linear-in-$\ell$ growth of the 1PN-coefficient in $\rho_{\ell m}$.  The function
$f_{\ell m}$ is then given by,

	\beq
	f_{\ell m} = \begin{cases}
	(\rho_{\ell m})^{\ell}, &m \text{ is even},\\
	(\rho_{\ell m})^{\ell} + f^{\rm S}_{\ell m}, &m \text{ is odd},
	\end{cases}
	\eeq
where $f^{\rm S}_{\ell m}$ denotes the spin-dependent contribution to the amplitude.
Similar resummations can be also be found in other EOB models, for example in \texttt{TEOBResumS} in 
Refs.~\cite{Nagar:2011aa,Nagar:2018zoe,Nagar:2019wds,Nagar:2020pcj,Nagar:2022fep}.

As described in Ref.~\cite{vandeMeent:2023ols}, $\texttt{SEOBNRv5}$ includes additional calibration of both the waveform modes and the RR force in the nonspinning sector
using 2GSF flux data \cite{Warburton:2021kwk}. One considers the expansion of the $\rho_{\ell m}$ in the terms of the symmetric mass-ratio,
	\beq
	\rho_{\ell m} = \rho^{(0)}_{\ell m} +  \nu \rho^{(1)}_{\ell m} + {\cal O}(\nu^{2}).
	\eeq
The linear in mass-ratio term, $\rho^{(1)}_{\ell m}$, is then modified by a correction term, $\Delta \rho^{(1)}_{\ell m}$, 
determined by fitting to SF results for the linear in mass-ratio term. After the publication of the \texttt{SEOBNRv5} model~\cite{vandeMeent:2023ols,Pompili:2023tna}, 
the authors of Ref.~\cite{Warburton:2021kwk} discovered an error in their data, which was used for the calibration. 
While redoing the calibration with the corrected SF data, an inconsistency was found with the way the SF-calibration was propagated to the waveform modes in Ref.~\cite{vandeMeent:2023ols}.  
When consistently computing the flux and waveform, the 2GSF calibration should yield distinct correction terms for the RR force \emph{and} the waveform modes. 
These findings will be addressed in detail in a forthcoming erratum.

In this work, we incorporate the corrected $\Delta \rho^{(1)}_{\ell m}$ flux terms, but not those for the waveform modes. 
This is justified because the amplitude corrections enter nominally at second post-adiabatic (2PA) order, 
where additional effects (e.g., waveform phase corrections) that are not currently included in \texttt{SEOBNRv5} also become relevant. Nevertheless, 
we find that when including the waveform-mode SF-corrections, the performance of the model is not significantly improved and the 
results for the unfaithfulness only marginally change for comparable-mass systems.

The correction terms $\Delta \rho^{(1)}_{\ell m}$ used in this work are given below, with differences from the original coefficients in 
Ref.~\cite{vandeMeent:2023ols} highlighted in \textbf{bold}:
\begin{subequations}
	\begin{align}
		\Delta \rho^{(1)}_{22} &= \bm{20.6} v_{\Omega}^{8} \bm{- 410} v_{\Omega}^{10},\\
		\Delta \rho^{(1)}_{21} &= 1.65 v_{\Omega}^{6} \bm{+ 24.3} v_{\Omega}^{8} + 80 v_{\Omega}^{10},\\
		\Delta \rho^{(1)}_{33} &= 12 v_{\Omega}^{8} \bm{- 222} v_{\Omega}^{10},\\
		\Delta \rho^{(1)}_{32} &= -\bm{0.25} v_{\Omega}^{6} - 6.5 v_{\Omega}^{8} + 98 v_{\Omega}^{10},\\
		\Delta \rho^{(1)}_{44} &= -3.56 v_{\Omega}^{6} + \bm{15.3} v_{\Omega}^{8} - 216 v_{\Omega}^{10},\\
		\Delta \rho^{(1)}_{43} &= -0.654 v_{\Omega}^{4} \bm{- 3.73} v_{\Omega}^{6} \bm{+ 18} v_{\Omega}^{10},\\
		\Delta \rho^{(1)}_{55} &= \bm{-2.608} v_{\Omega}^{4} \bm{+ 1.13} v_{\Omega}^{6} \bm{- 35.1} v_{\Omega}^{10}.
	\end{align}
\end{subequations}

The non-quasicircular corrections, $N_{\ell m}(t)$, for \texttt{SEOBNRv5} have the following form:
\begin{align}
N_{\ell m} &= \left[1+ \frac{p_{r_{\star}}^2}{(r\Omega)^2}
	\left( a_1^{h_{\ell m}} + \frac{a_2^{h_{\ell m}}}{r} + \frac{a_3^{h_{\ell m}}}{r^{3/2}} \right)
	\right]
	\nonumber \\
	&\qquad\times
	\exp\left[i \left(b_1^{h_{\ell m}} \frac{p_{r_{\star}}}{r\Omega} + b_2^{h_{\ell m}} \frac{p_{r_{\star}}^3}{r\Omega} \right)\right],
\end{align}
with the constants $a_{i}^{h_{\ell m}}$ and $b_{i}^{h_{\ell m}}$ ($i \in [0,3]$) fixed such that the EOB modes match the equivalent modes from NR at the point 
$t_{\rm match}$.
It is important to state here, that unlike the \texttt{TEOBResumS} models and earlier iterations of \texttt{SEOBNR}, the NQCs in \texttt{SEOBNRv5} do not enter the RR 
forces for the model \cite{vandeMeent:2023ols, Pompili:2023tna}.

The merger-ringdown EOB gravitational waveform modes, $h_{\ell m}^{\rm{merger-RD}}(t)$, are independent of the inspiral-plunge
modes but they are smoothly connected to the inspiral-merger waveform by requiring continuity and differentiability of the waveform at 
merger. The merger-ringdown modes are based upon a phenomenological ansatz given explicitly in 
\cite{Damour:2014yha, Bohe:2016gbl, Cotesta:2018fcv,Pompili:2023tna}, which is then supplemented by NR simulations as well as TML waveforms 
\cite{Barausse:2011kb,Taracchini:2014zpa}.
Once again, we direct the reader to Ref.~\cite{Pompili:2023tna}, for a detailed overview of their construction.
It suffices to state that the merger-ringdown modes incorporated in the model presented herein are identical to those utilized in \texttt{SEOBNRv5}.

\subsection{SEOBNRv5 equations of motion and radiation-reaction force}

The EOB equations of motion are given by
\begin{align}
	\label{eq:eob_eom}
	\begin{aligned}
		\diff{r}{t} &= \xi \frac{\partial H}{\partial p_{r_{\star}}}, \quad
		&\diff{{p}_{r_{\star}}}{t} &= -\xi \frac{\partial H}{\partial r} + \frac{p_{r_{\star}}}{p_\varphi} {\cal F}_\varphi, \\
		\diff{\varphi}{t} &= \frac{\partial H}{\partial p_\varphi}, \quad
		&\diff{{p}_\varphi}{t} &= {\cal F}_\varphi.
	\end{aligned}
\end{align}
Here we use the tortoise-coordinate radial momentum $p_{r_{\star}}$ as opposed to the canonical radial momentum, 
$p_{r}$.  
This choice enhances the stability of the equations of motion throughout the plunge phase and near the merger, as 
documented in Refs.~\cite{Damour:2007xr, Pan:2009wj}.
In the usual Hamiltonian prescription in \eqn{h_eff_generic} for nonspinning binaries, 
the tortoise-coordinate $\rstar$ is defined as 
in~\cite{Damour:2008gu} by
\begin{equation}
	\diff{\rstar}{r} = \frac{1}{\xi(r)}, \qquad
	\xi(r) := A(r)\sqrt{\bar{D}(r)},
\end{equation}
with the conjugate momentum $p_{\rstar}$ given by
\beq
	p_{\rstar} =  p_{r} \,\xi(r).
	\label{eq:prstar}
\eeq
These definitions simplify in the PS-gauge, since $A(r)$ and $\bar{D}(r)$ are taken to be in the Schwarzschild
TML, hence
\beq
	\xi_{\rm PS}(r) = f(r) = 1 - \frac{2M}{r}.
\eeq

The RR force in this setup, ${\cal F}_\varphi$, is found from the summation of the EOB gravitational waveform modes:
\beq
	{\cal F}_{\varphi} = -\frac{{\cal F}^{\rm EOB}}{M\Omega} = -\frac{1}{M\Omega} \sum_{\ell=2}^{8} \sum_{m=1}^{\ell} 	
	{\cal F}_{\ell m}^{\rm EOB},
	\label{eq:radiation_reaction_force}
\eeq
where the $(\ell, m)$-contributions are found directly from the inspiral-plunge waveform modes in \eqn{inspiral_plunge_form}
such that
\beq
	{\cal F}_{\ell m}^{\rm EOB} = d^{2}_{L} \frac{(m M \Omega)^{2}}{8\pi} \abs{h_{\ell m}^{{\rm F}}}^{2}.
	\label{eq:radiation_reaction_force_modes}
\eeq

\section{Calibration to numerical-relativity waveforms}
\label{sec:calibration_with_nr}
We shall now explore the calibration pipeline for our model with NR simulations.
For the full \texttt{SEOBNRv5} model, there are three calibration parameters: 
$(a_{6}, d_{\rm SO}, \Delta t_{\rm NR})$, which are each a function of the binary
parameters $\bm{\lambda}$, and are designed to improve the agreement with binary BH simulations.
The $a_{6}$ parameter replaces the partially known 5PN-coefficient ($u^{6}$-coefficient) in
the DJS-gauge $A(u)$-potential, whilst the second parameter, $d_{\rm SO}$,
is a 4.5PN spin-orbit parameter \cite{Pompili:2023tna}.
Since our model is in the nonspinning regime and our primary potential is informed through SF information rather
than purely a PN expansion, we do not explore calibration parameters akin to $a_{6}$ and $d_{\rm SO}$ in this work.
In this section our focus will therefore be on the sole calibration parameter $\Delta t_{\rm NR}$.

Introduced in \eqn{t_22_peak}, $\Delta t_{\rm NR}$ is a parameter that determines the time shift between the ISCO, computed
from the final mass and spin of the remnant object~\cite{Jimenez-Forteza:2016oae, Hofmann:2016yih} 
and the peak of the (2,2)-mode amplitude.
The idea is to allow $\Delta t_{\rm NR}$ to be a freely varying parameter that is determined by minimizing the discrepency
of the waveforms' peak time, between EOB and NR, at the end of the inspiral-plunge stage (i.e. $t < t_{\rm match}$). 
To do this we calibrate our model in a similar manner to \texttt{SEOBNRv5} as outlined in Ref.~\cite{Pompili:2023tna}, to a set of nonspinning
NR waveforms all generated using the pseudo-Spectral Einstein code (SpEC) from the Simulating eXtreme Spacetimes (\texttt{SXS}) collaboration
\cite{SXS:catalog, Boyle:2019kee, Chu:2015kft, Blackman:2015pia, Hemberger:2013hsa, Lovelace:2014twa, Bohe:2016gbl, Blackman:2017dfb, Lovelace:2016uwp, Varma:2018mmi, 
Varma:2019csw, Mroue:2013xna, Yoo:2022erv}.
This publically avaliable subset of nonspinning waveforms spans mass ratios in the range $1 < q < 20$ and forms
part of the larger set used for the full calibration procedure described in Ref.~\cite{Pompili:2023tna}, which 
covers a broader region of parameter space.
They are the same set of NR simulations used in \texttt{SEOBNRv5} and \texttt{SEOBNRv5-PM}.
The specific simulations used in this work are listed in Table~\ref{tbl:sxs_sims} of Appendix~\ref{sec:nr_sims}.

A crucial aspect required to be able to calibrate an EOB waveform to an NR waveform is an appropiate measure to establish 
the difference between two waveforms.
Therefore we define the noise-weighted inner product, $\left(h_1 \mid h_2\right)$, between two different waveforms, 
$h_{1}(t)$ and $h_{2}(t)$ \cite{Finn:1992xs, Sathyaprakash:1991mt}:
\beq
	\left(h_{1} \mid h_{2}\right) := 4 \operatorname{Re} \left [\int_{f_{l}}^{f_{h}} \frac{\tilde{h}_{1}(f) 
	\tilde{h}_{2}^*(f)}
	{S_{n}(f)} {\rm d} f \right ].
	\label{eq:inner_product}
\eeq
Here, $S_{n}(f)$ is the one-sided power spectral density of the detector noise of advanced LIGO, see 
Ref.~\cite{Barsotti:2018}, and $\tilde{h}_{i}$ denotes the Fourier transform of the waveform $h_{i}$.
We also fix $f_{h} = 2048~{\rm Hz}$.
The second limit of integration, $f_{l}$, is dependent on the type of waveforms that enter the inner product.
If the two waveforms are in band, such as when one is comparing two different EOB waveform models, $f_{l} = 10~{\rm Hz}$.
However, when one is comparing to NR waveforms, we set $f_{h}$ to have a buffer factor such that $f_{l} = 1.35 f_{\rm start}$,
where $f_{\rm start}$ is the peak of the NR waveform in the frequency domain.
The inclusion of the buffer factor is to ensure that any spurious features that enter into the inner product due to the Fourier transform
of the time-domain NR waveform are excluded from the calculation of the inner product.
This inner product is then used to calculate the \emph{faithfulness}, between two waveforms
\beq
	\left\langle h_{1} \mid h_{2} \right\rangle =
	\max_{\delta\phi,\delta{}t} \frac{( h_{1}(\delta \phi, \delta t)\vert h_{2} )}
	{\sqrt{(h_{1} \vert h_{2})( h_{1}\vert h_{2} )}},
	\label{eq:faithfulness}
\eeq
which is the overlap over the two waveforms maximized over the time and phase shift, $\delta t$ and $\delta \phi$,
respectively.
One can then use the faithfulness to define the \emph{unfaithfulness}, otherwise known as \emph{mismatch}, between an EOB waveform
an NR waveform, denoted $h_{\rm EOB}$ and $h_{\rm NR}$ respectively:
\beq
	{\cal M}(\bm{\theta}) := 1 - \left\langle h_{\rm EOB}(\bm{\lambda}; \bm{\theta}) \mid h_{\rm NR}(\bm{\lambda}) \right\rangle.
	\label{eq:unfaithfulness}
\eeq
Here, we have introduced $\bm{\theta}$ as the generic set of calibration parameters.  
Unlike in Ref.~\cite{Pompili:2023tna}, in this work we are only exploring the use of a single calibration parameter such that 
$\bm{\theta} := \Delta t_{\rm NR}$.

\subsection{Calibration of $\Delta t_{\rm NR}$}
\label{sec:delta_t_22_calibration}
In calibrating our $\Delta t_{\rm NR}$\footnote{Note in
Ref.~\cite{Pompili:2023tna}, what we refer to as $\Delta t_{\rm NR}$ was labelled as $\Delta t^{22}_{\rm ISCO}$.} parameter, 
we adopt the same strategy as Ref.~\cite{Pompili:2023tna} that was used in 
the calibration of the \texttt{SEOBNRv5} model, which itself was based on the methods used in calibrating the previous version 
in this family of EOB models, \texttt{SEOBNRv4} \cite{PhysRevD.95.044028, PhysRevD.95.044028, PhysRevD.98.084028}.
Our aim is to fix the calibration parameter by creating a fit over the binary parameter space.
Since we are only considering nonspinning, quasicircular binary systems, the fit shall only be dependent on the symmetric mass ratio 
and is obtained by sampling over each nonspinning quasicircular NR simulation.
One samples over a 1-dimensional grid to obtain a posterior distribution the $\Delta t_{\rm NR}$ parameter for each NR simulation from a 
uniform prior,
before taking a direct fit of the maximum-likelihood point of the posterior distributions.
Note that, unlike the calibration procedure described in Ref.~\cite{Pompili:2023tna}, the calibration procedure here does 
not require the use of a Markov-chain Monte Carlo (MCMC) method and, consequently, does not require a nested sampling 
algorithm, as our calibration parameter space is one-dimensional rather than multi-dimensional.
This would change if one was considering a more complete calibration procedure with extra PN and spinning calibration
parameters as is the case with \texttt{SEOBNRv5}.

We define the maximum-likelihood as
\begin{multline}
	P(h_{\rm NR} \vert \bm{\theta}) \propto 
	\exp \bigg[ -\frac{1}{2} \left(\frac{{\cal M}_{\rm max}({\bm \theta})}{\sigma_{\cal M}}\right)^{2} \\
	-\frac{1}{2} \left( \frac{\delta t_{\rm merger}({\bm \theta})}{\sigma_{t}} \right)^{2} \bigg],
	\label{eq:max_likelihood}
\end{multline}
where ${\cal M}_{\rm max}$ is the maximum unfaithfulness over total mass between the NR and EOB waveforms, see \eqn{unfaithfulness}, and
$\delta t_{\rm merger}$ is the difference in the merger\footnote{In this context the merger time is defined as the peak of the amplitude
of the $(2,2)$-mode.} times of the NR and EOB waveforms.
We choose the parameters, $\sigma_{\cal M}$ and $\sigma_{t}$, to be $10^{-3}$ and $5M$ respectively, which are the same values used in the
calibration of \texttt{SEOBNRv5} \cite{Pompili:2023tna}.
This maximum-likelihood is the measure on which we perform the basis of our fitting procedure.

During the sampling routine, we assume uniform priors for the calibration parameter and vary over the range 
$\Delta t_{\rm NR} \in [-100, 40]$ to obtain posterior distributions for $P(h_{\rm NR} \vert \bm{\theta})$.
One difficulty that arises for some values of posterior distributions is multimodality.  
That is, secondary modes can appear for certain values of the binary parameters, ${\bm \theta}({\bm \lambda})$.
These modes would spoil a potential fitting of the calibration parameter and so to ensure a more uniform fit, we choose just one mode from 
each calibration posterior, considering continuity.  
In practice, one can directly discard samples from the original posterior, but here we choose to separate the distributions through fitting the posterior
using a \emph{Gaussian mixture model} (GMM) through the \texttt{sklearn.mixture} package \cite{scikit-learn}.
This model assumes the distribution is made of a mixture of Gaussian distributions, each with its own mean and covariance. 
Therefore it is particularly effective for capturing the presence of subpopulations within an overall population, without requiring prior knowledge of which 
subpopulation an observation belongs to.
The model parameters are estimated using the Expectation-Maximization (EM) algorithm \cite{scikit-learn}, which iteratively assigns a ``soft" membership of each data point to a 
particular Gaussian component and updates the Gaussian parameters accordingly. 
The process continues until convergence, yielding a robust separation of the posterior into distinct modes. 
Once this analysis is complete, we identify bimodal posteriors and retain only the component corresponding to the dominant mode, extracting its mean, variance, 
and maximum-likelihood estimate.

\begin{figure}[h]
	\includegraphics[width=\linewidth]{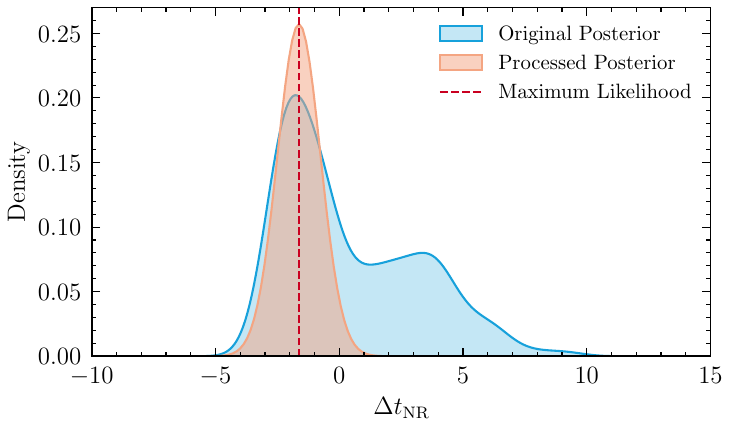}
	\caption{An example of a multimodal posterior for the calibration parameter $\Delta t_{\rm NR}$ for 
	\texttt{SXS:BBH:2425}
	$(q = 2.0, \chi_{1} = \chi_{2} = 0.0)$.
	The blue posterior is the initial result we obtain from our sampling over the range of $\Delta t_{\rm NR}$ 
	values and exhibits
	two distinct regions or \emph{modes} in which the values cluster around.  
	The second, orange, posterior is a processed posterior where we have removed the mode corresponding to the 
	postive value of $\Delta t_{\rm NR}$ as this would lead to irregular fit of the calibration parameter.}
	\label{fig:multimodal_distributions}
\end{figure}

An example multimodal posterior is given in \fig{multimodal_distributions}.
We find a correlation between the appearance of multimodal posteriors and more comparable mass-ratio simulations.
These correlations manifest themselves for mass-ratios $\nu \gtrsim 0.14$ $(q \lesssim 5)$, the region of the parameter space where we expect
the small-mass ratio approximation of the SF expansion to be at its least accurate.

\begin{figure}[h]
	\includegraphics[width=\linewidth]{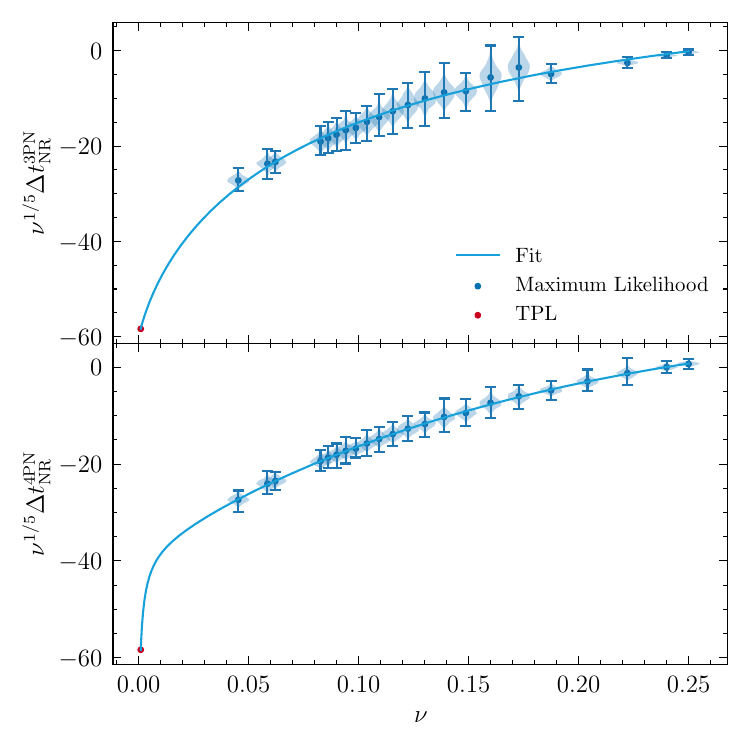}
	\caption{Fits for the $\Delta t_{\rm NR}$ calibration parameter for the
	3PN and 4PN \texttt{\texttt{SEOBNR-GSF}} models.  
	One obtains the calibration parameter as a function of the mass-ratio $\nu$ from
	a least squares fit of the maximum likelihood of the posteriors (blue shaded violins) 
	for a set of nonspinning NR simulations and TML values.
	As done for \texttt{SEOBNRv5} in Ref.~\cite{Pompili:2023tna}, we adopt the rescaling
	by $\nu^{1/5}$, which substantially improves the agreement in the test-particle limit (TPL) 
	where $\nu \rightarrow 0$.}
	\label{fig:delta_t_22_fits}
\end{figure}

After processing the posteriors when appropiate, we also leverage information of the conservative dynamics in the TML to extrapolate our fit to the $\nu \rightarrow 0$ limit
using Ref.~\cite{Taracchini:2014zpa} in the same way as done in Ref.~\cite{Pompili:2023tna}.
The resultant posteriors and TML value are then fitted directly using a least-squares-fitting procedure and 
a rational function.  In our investigations, we find that a different choice of rational function works better for each variant
of the model with 3PN and 4PN corrections.  For the model with 3PN corrections, the same rational function
as Ref.~\cite{Pompili:2023tna} produces a more natural fit, given by
\beq
	\Delta t_{\rm NR}^{3{\rm PN}} = (a^{3}_{0} + a^{3}_{1} \,\nu + a^{3}_{2} \,\nu^2 + a^{3}_{3} \,\nu^3 )
	\,\nu^{-1/5 + a^{3}_{4}\, \nu},
	\label{eq:delta_t_NR_3pn_fit}
\eeq
where $a_{i}\, (i \in [0, 4])$ are constants determined by the fitting procedure.
Note with this we retain the same functional form as Ref.~\cite{Pompili:2023tna} including the expected test mass scaling of 
$t_{\rm peak} - t_{\rm ISCO}$ with the $\nu^{-1/5}$ factor \cite{Buonanno:2000ef}.
For the model with the 4PN corrections, however, we find a more reasonable fit using a fitting function inspired by a 
\emph{post-leading-transition expansion} in Ref.~\cite{Kuchler:2024esj},
\beq
	\Delta t_{\rm NR}^{4{\rm PN}} = (a^{4}_{0} + a^{4}_{1} \,\nu^{1/5} + a^{4}_{2} \,\nu^{2/5} + a^{4}_{3} \,\nu^{3/5} 
	+ a^{4}_{4}\, \nu^{4/5}) \,\nu^{-1/5},
	\label{eq:delta_t_NR_4pn_fit}
\eeq
which better fits the resulting trend of the posteriors.
The resultant fits for the 3PN and 4PN corrected versions of the \texttt{\texttt{SEOBNR-GSF}} models are shown in 
\fig{delta_t_22_fits}, which we refer to as $\Delta t_{\rm NR}^{3{\rm PN}}$ and $\Delta t_{\rm NR}^{4{\rm PN}}$
respectively.
Explicitly, the fitting coefficients for the calibration parameter for both models read,
\begin{align}
	a^{3}_{0} &=  6.06749591, \nn\\
	a^{3}_{1} &=  -58.53430766, \nn\\
	a^{3}_{2} &=  -274.89832051, \nn\\
	a^{3}_{3} &=  2552.05275465, \nn\\
	a^{3}_{4} &= -1981.68604241, \nn\\
	\nn\\
	a^{4}_{0} &=  -205.15155028, \nn\\
	a^{4}_{1} &=  1188.35994636, \nn\\
	a^{4}_{2} &=  -3307.97319917, \nn\\
	a^{4}_{3} &=  4191.43478856, \nn\\
	a^{4}_{4} &= -1876.6265344.
	\label{eq:delta_t_22_fitting_coefficients}
\end{align}

\begin{figure}[h]
	\includegraphics[width=\linewidth]{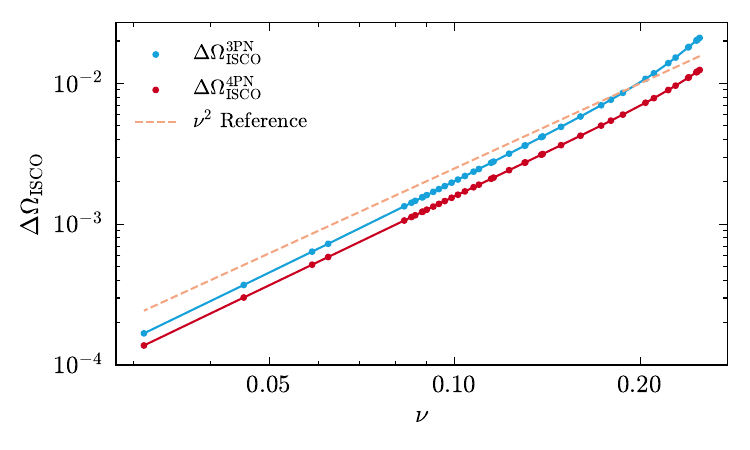}
	\caption{Residuals of the ISCO frequency shift relative to the first-order GSF expansion. 
	As expected, both the 3PN and 4PN versions of the \texttt{SEOBNR-GSF} model exhibit the correct TML 
	behaviour: the residuals scale as $\nu^2$, consistent with the next-to-leading order in the GSF expansion as 
	$\nu \rightarrow 0$. The curves show residuals computed from ISCO frequency shift values extracted from the 
	publicly available nonspinning NR simulations in the \texttt{SXS} catalogue.}
	\label{fig:delta_omega_isco}
\end{figure}

As mentioned previously, we do not seek a calibration parameter akin to the $a_{6}$ parameter employed in the
\texttt{SEOBNRv5} model since our model is primarily informed from GSF information.
But, the $a_{6}$ calibration parameter in the TML is tuned using data for the frequency shift of the ISCO calculated using GSF results 
\cite{Barack:2010tm, Isoyama:2014mja, Akcay:2012ea} such that
\begin{align}
	\begin{aligned}
	M \Omega_{\rm{ISCO}}^{\rm{1SF}} &= 6^{-3/2} (1 + C_{\Omega} / q), \\
	C_{\Omega} &= 1.25101539 \pm 4\times 10^{-8}.
	\label{eq:isco_shift_gsf}
	\end{aligned}
\end{align}
Therefore, one would expect a GSF-informed model would capture this ISCO frequency shift in the TML.
In the EOB formalism, the ISCO can be computed directly from the EOB Hamiltonian by finding the point in phase space 
$(r, p_{\varphi})$ where the first and second derivatives of the Hamiltonian vanish when $p_{r} = 0$:
\beq
	\psdiff{H}{r}\bigg\vert_{p_{r} = 0} = \pdiff{H}{r}\bigg\vert_{p_{r} = 0} = 0.
	\label{eq:isco_equation}
\eeq
We find that both of our SF models produce agreement with the $\Omega_{\rm{ISCO}}^{\rm{1SF}}$ value in the TML
with a relative error of $\lesssim 0.4\%$, whilst \texttt{SEOBNRv5}, without calibration of the $a_{6}$ parameter to 
account for the ISCO shift, yields a value with a relative error of $\sim 6.7\%$.
Furthermore, one can assess the scaling of the residual of the frequency shift calculated from the EOB model and the
GSF expansion of \eqn{isco_shift_gsf}.
As one can see in \fig{delta_omega_isco}, we obtain the expected $\nu^{2}$-scaling in the limit $\nu \rightarrow 0$
for the residual of the frequency shift, denoted $\Delta\Omega_{\mathrm{ISCO}}$, for both flavours of \texttt{\texttt{SEOBNR-GSF}}
model.
It should also be noted, although we have not included the result in the plot for visual simplicity, that the \texttt{SEOBNRv5} model
without the $a_{6}$ calibration does not reproduce the same behaviour in the TML.

\section{Results}
\label{sec:results}
\subsection{Binding energy}
\begin{figure*}[t!]
	\includegraphics[width=\linewidth]{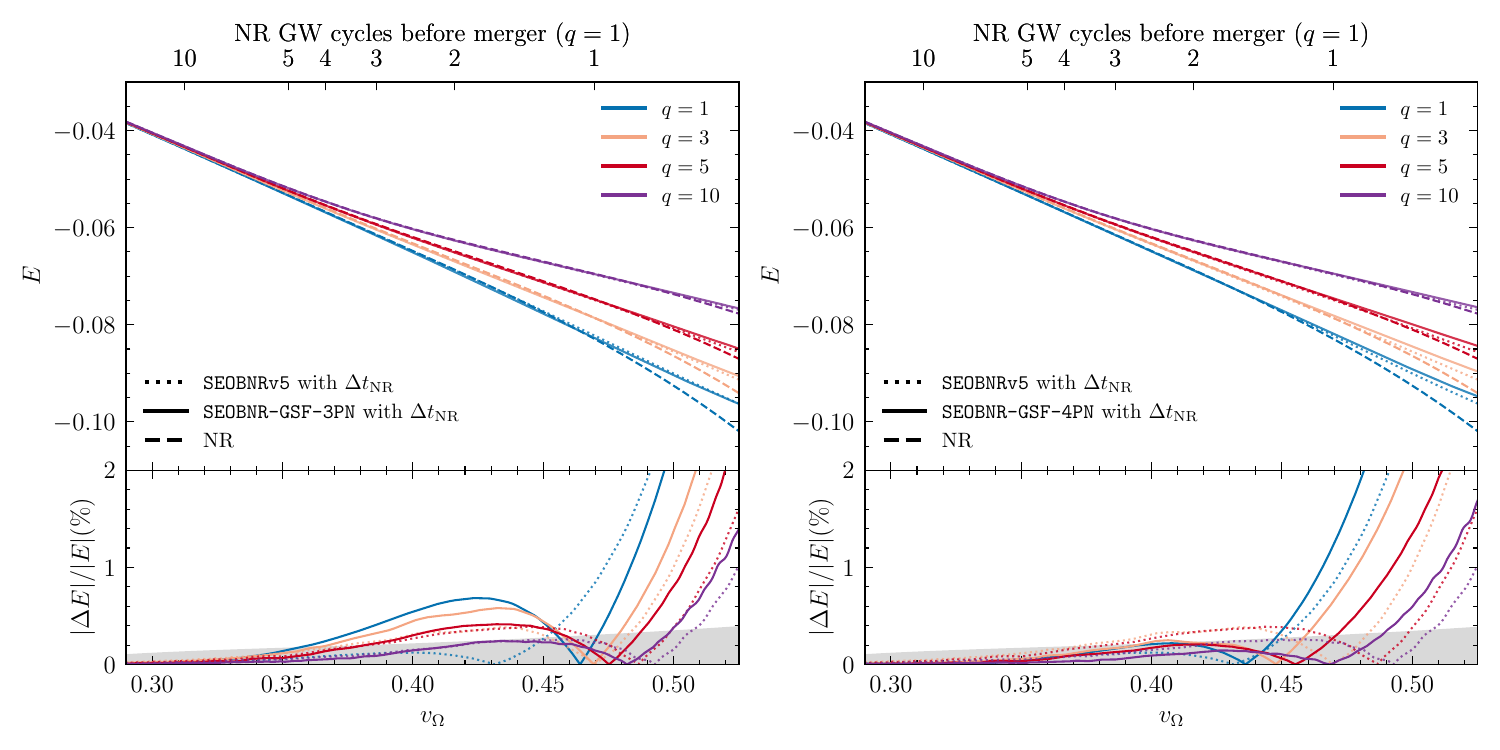}
	\caption{The nonspinning binding energy as a function of the orbital velocity, $v_{\Omega} = (M \Omega)^{1/3}$,
		is presented for the \texttt{SEOBNRv5} model with a calibrated $\Delta t_{\rm NR}$ parameter and the 
		\texttt{\texttt{SEOBNR-GSF}} model incorporating 3PN corrections (left plot) and 4PN corrections (right plot), 
		both with a calibrated $\Delta t_{\rm NR}$ parameter for a range of mass-ratios $q \in [1, 10]$.
		The upper panels show the binding energy $(E = E_{\rm bind})$ as a function of $v_{\Omega}$ along the orbit, 
		whilst the lower panels show the fractional difference 
		(where $\Delta E = E^{X}_{\rm bind} - E^{\rm NR}_{\rm bind}$) 
		between the EOB binding enegies for the various models and the NR simulations across different mass-ratios.  
		The shaded region indicates an estimate of the
		uncertainty of the NR simulation for $q = 1$, as provided in Ref.~\cite{Ossokine:2017dge}, and similarly for 4PN, 
		and \texttt{SEOBNRv5} with $\Delta t_{\rm NR}$.}
	\label{fig:binding_energy}
\end{figure*}

The first assessment of the performance of our model is to compare the binding energy of the binary system.
This serves as a complimentary assessment to the computing of the unfaithfulness with NR waveforms, giving us an insight into how the 
information from the two-body dynamics is encoded in the resummation of the conservative dynamics that forms our model.
A similar comparison was completed in Ref.~\cite{Antonelli:2019fmq}.
In \fig{binding_energy} we present the fractional difference between the 
EOB binding energy, $E^{\rm EOB}_{\rm bind}$, and the binding energy obtained from NR simulations for four 
different mass ratios, using the 3PN and 4PN models, respectively, as well as the \texttt{SEOBNRv5} model 
calibrated solely with the $\Delta t_{\rm NR}$ parameter.
For the NR simulations, we utilize the binding energy obtained in Ref.~\cite{Ossokine:2017dge}.
The binding energies in both panels of \fig{binding_energy} 
are plotted as a function of the quasi-circular velocity parameter,
$v_{\Omega}$, where the orbital frequency $\Omega = \Omega_{\rm NR}$ for the NR simulations 
and $\Omega_{\rm EOB}$ is the orbital frequency for the EOB models.
In particular, $\Omega_{\rm EOB}$ is obtained directly from 
$\Omega_{\rm EOB} = \partial H / \partial \varphi$ to ensure there is no gauge ambiguity between the
EOB models and the definition used in NR simulations \cite{Ossokine:2017dge}.
It should emphasised that the EOB binding energy here is evaluated along the orbital dynamics.

We see very good agreement between our new EOB models and NR up to a few cycles before merger, with markedly
better agreement for our 4PN model, which remains within the uncertainty of our NR simulations up to 
1 cycle before merger.  This is a similar level of accuracy that we see with \texttt{SEOBNRv5}, as is
evident from both figures.  Unsurprisingly, this shows that the \texttt{SEOBNR-GSF}
models with 4PN-corrections to perform better than the model with only up to 3PN corrections.
Therefore in our waveform performance comparisons in the next section, we only consider comparisons
with the \texttt{SEOBNR-GSF} model with 4PN corrections to avoid unnecessary comparisons.
Henceforth, any plot of \texttt{SEOBNR-GSF} will be referring to the model with 4PN corrections unless
explicitly stated otherwise. 

\subsection{Comparison to \texttt{1PAT1} waveforms}
\begin{figure*}[t]
	\includegraphics[width=0.9\linewidth]{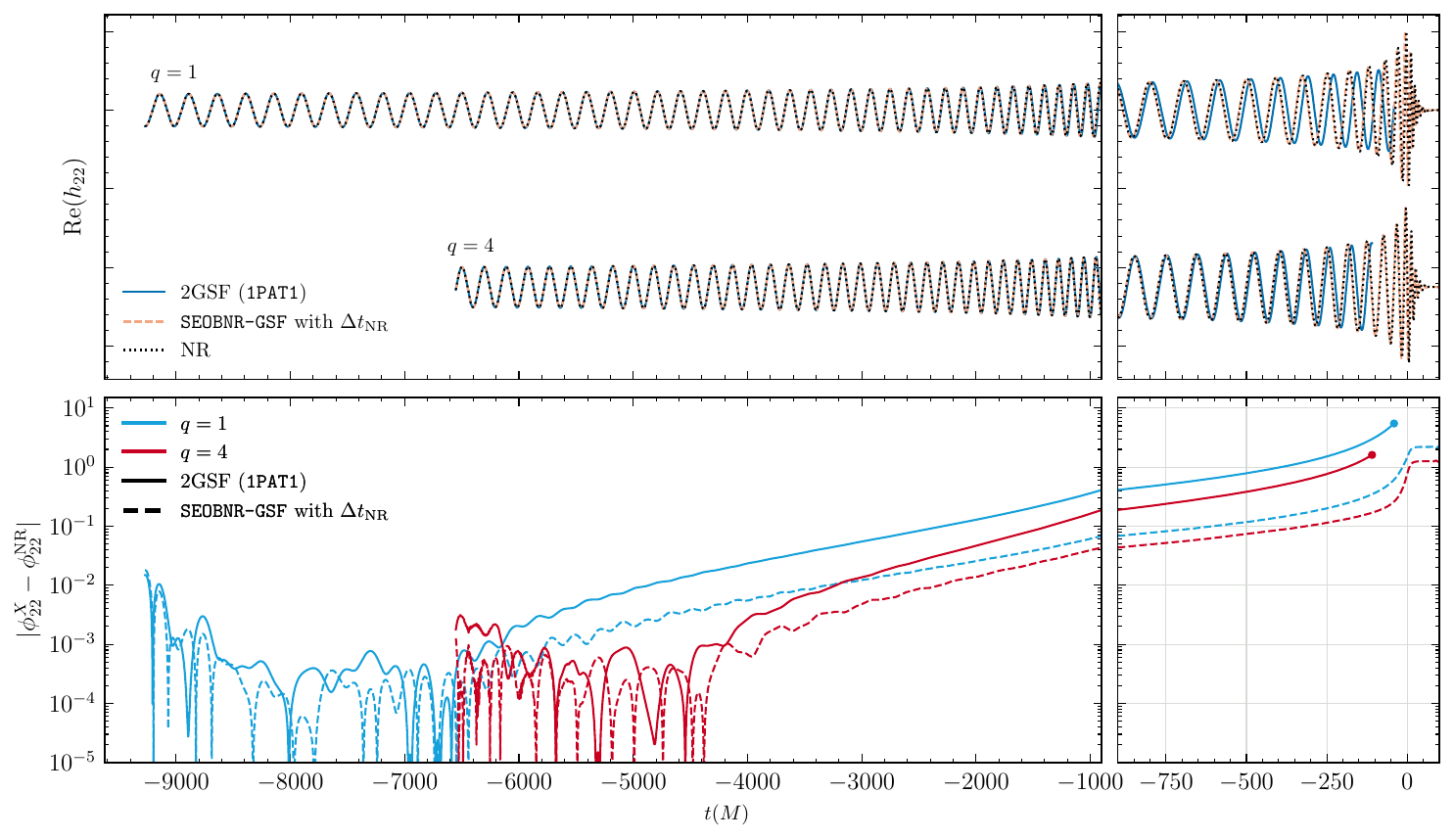}
	\caption{Top panel: A comparison of the (2,2) waveform modes of the \texttt{1PAT1} model, the
	\texttt{SEOBNR-GSF} model and NR simulations at mass-ratios $q = 1$ and $q = 4$.  
	The waveforms are aligned at early times using the procedure originally outlined in 
	Ref.~\cite{Pan:2011gk}.  
	Bottom panel: The phase difference between the two models and the NR simulations for the waveforms
	plotted in the top panel.  In both panels, we magnify last $-900M$ before merger (peak of the 
	$(2, 2)$-mode amplitude) and end the \texttt{1PAT1} waveforms at $v_{\Omega}$.
	One observes at these small mass-ratios, the \texttt{SEOBNR-GSF} model has a markedly smaller phase 
	difference when compared to \texttt{1PAT1} model, particularly towards the late inspiral towards 
	$v^{\rm break}_{\Omega}$, where the multiscale expansion breaks down.  Furthermore, the \texttt{SEOBNR-GSF} model
	includes all stages of the binary coalescence through inspiral, merger and ringdown.}
	\label{fig:waveform_comparison_low_q}
\end{figure*}

\begin{figure*}[t]
	\includegraphics[width=0.9\linewidth]{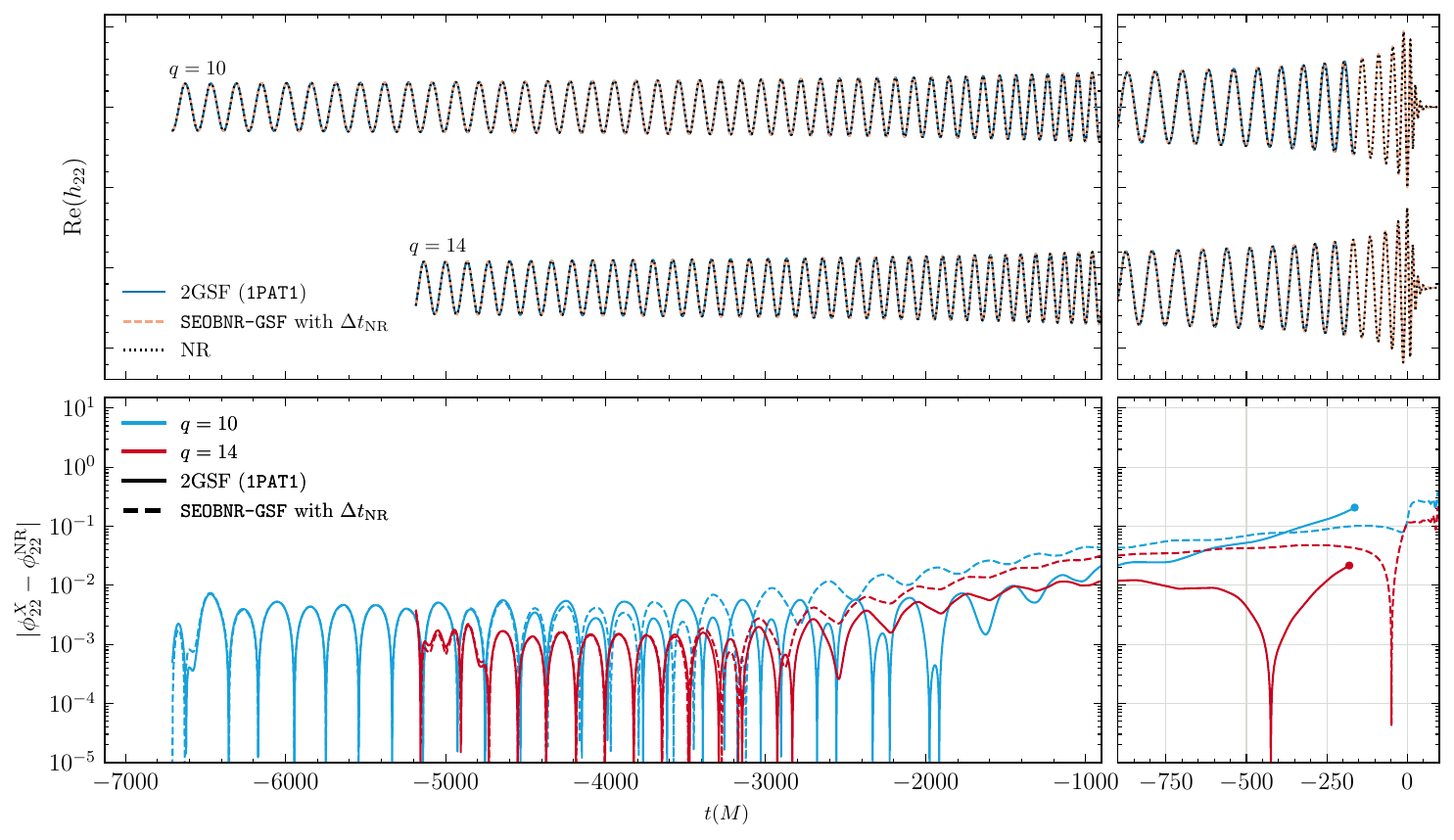}
	\caption{Top panel: A comparison of the (2,2) waveform modes of the \texttt{1PAT1} model, the
	\texttt{SEOBNR-GSF} model and NR simulations at mass-ratios $q = 10$ and $q = 15$.  
	Bottom panel: The phase difference between the two models and the NR simulations for the waveforms
	plotted in the top panel.
	For these higher mass-ratios, the \texttt{SEOBNR-GSF} model has a comparable phase 
	difference when compared to \texttt{1PAT1} model, with the main difference being the \texttt{SEOBNR-GSF}
	model does not break down before the merger.}
	\label{fig:waveform_comparison_high_q}
\end{figure*}

The crucial measure of our model's performance lies in the quality of the waveforms it generates.
In Figs.~\ref{fig:waveform_comparison_low_q} and \ref{fig:waveform_comparison_high_q}, 
we present a comparison of the $(2,2)$-mode of the \texttt{SEOBNR-GSF} waveforms with those 
generated directly from 2GSF calculations, specifically the \texttt{1PAT1} variant \cite{Wardell:2021fyy}, 
as well as with NR waveforms from the \texttt{SXS} catalogue. This comparison is conducted for both 
low mass ratios $(q = 1, 4)$ in \fig{waveform_comparison_low_q} and high mass ratios $(q = 10, 15)$
in \fig{waveform_comparison_high_q}.  

A detailed overview of the \texttt{1PAT1} GSF waveform model,
and the approximations made therein, is found within Sec. II of Ref.~\cite{Albertini:2022rfe}. 
We briefly outline the important details here for our subsequent discussion.

The \texttt{1PAT1} model is based on a \emph{multiscale expansion} of Einstein's field equations, 
where the metric perturbation is expanded to separate the slow evolution of the binary's orbital 
frequency, $\Omega$, from the rapidly evolving phase, $\varphi$. In this framework, the slow 
evolution occurs on the radiation-reaction timescale, $t_{\rm RR} \sim (q \Omega)^{-1}$, 
while the orbital phase evolves on the much shorter orbital timescale, 
$t_{\rm orb} \sim (\Omega)^{-1}$. These timescales are treated independently in the analysis.

This decomposition results in a set of Fourier-domain partial differential equations (PDEs) 
for the amplitudes of the metric perturbation at a fixed $\Omega$ and a separate set of 
ordinary differential equations (ODEs) governing the evolution of the binary's mechanical 
parameters, $\Omega$ and $\varphi$. The Fourier-domain PDEs are further simplified through 
a multipole decomposition, and by imposing the Lorenz-gauge condition \cite{Barack:2007tm}.
Solving these field equations as an offline step has been the main challenge in second-order 
SF calculations.
This has prompted a series of works exploring the groundbreaking calculations required for this 
step \cite{Pound:2009sm, Pound:2014xva, Pound:2015wva, Miller:2020bft, Miller:2023ers, Wardell:2015ada}.

The discussion in Ref.~\cite{Albertini:2022rfe} also highlights the domain of validity for the 
\texttt{1PAT1} model, which are valid only during the inspiral and break down towards the ISCO.
The reason for this breakdown lies within the assumptions of the multiscale expansion.
Implicitly, the expansion we have described assumes the secondary BH in the binary configuration
follows a quasicircular inspiral, whereby the variation of the orbital frequency is such that
$\dot{\Omega} \sim {\cal O} (q)$.  But this assumption breaks down close to the ISCO in such a way
for one encounters unphysical divergences in the multiscale expansion of the field equations, leading to
inaccuracies in the solutions to the field equations and quantities calculated therefrom, including the
second-order fluxes and therefore the waveforms from the \texttt{1PAT1} model.
To quantify where this limit of the model occurs, Ref.~\cite{Albertini:2022rfe} 
(and subsequently reported in Ref.~\cite{vandeMeent:2023ols}) 
provided an estimate for where multiscale expansion breaks down,
\begin{align}
	v^{\rm break}_{\Omega} &= v^{\rm ISCO}_{\Omega} - 0.052\nu^{1/4},\\
	v^{\rm ISCO}_{\Omega} &= \frac{1}{\sqrt{6}} \simeq 0.408.
	\label{eq:1PAT1_break}
\end{align}
We indicate this breakdown in both \fig{waveform_comparison_low_q} and \fig{waveform_comparison_high_q} 
with dots at the end of the phase difference in the bottom panels.

Immediately, in the top panel of each figure, one can distinguish the main difference between 
\texttt{SEOBNR-GSF} and \texttt{1PAT1} is that our EOB model includes the complete binary coalescence 
through inspiral, merger and ringdown whilst the 2GSF model breaks down towards merger as expected.
Remarkably, however, in the early inspiral regime, the waveforms are indiscernible from each other, 
it is only through looking at the phase difference in the bottom panels that we are able to separate
them.  The dephasing between the two models and NR simulations at lower mass-ratios is more 
pronouced in later stages of the inspiral, yet the phase difference between the \texttt{1PAT1}
waveforms and NR still remains $\lesssim 0.3$ radians even $\sim 900M$ before merger.  In the same regime,
the \texttt{SEOBNR-GSF} waveforms perform better for the same mass-ratios where the dephasing 
between the waveforms and NR are still below $\sim 0.1$ radians.  It is important that a similar study was 
conducted for similar mass-ratios for \texttt{SEOBNRv5} in Ref.~\cite{vandeMeent:2023ols}, 
which showed a comparable level of dephasing.

For the more intermediate mass-ratio comparison in \fig{waveform_comparison_high_q}, we see the 
\texttt{1PAT1} model remain at a more comparable level of dephasing to \texttt{SEOBNR-GSF}.
For $q = 10$ and $q = 15$, one sees that the \texttt{1PAT1} model only dephase by $\sim 0.5$
radians by the estimated break down at $v_{\Omega} \sim 150M$ before merger, with
\texttt{SEOBNR-GSF} at a similar level of dephasing.  The clear advantage of \texttt{SEOBNR-GSF} for these 
intermediate mass ratios is that the model maintains this level of dephasing while continuing through 
the latter stages of the inspiral, up to and including the merger and post-merger regime.

%\clearpage
\subsection{Inspiral-merger-ringdown comparison}

Since \texttt{SEOBNR-GSF} generates complete IMR waveforms, rather than only 
covering the inspiral phase, one can perform a more comprehensive assessment of the model’s performance by 
calculating the mismatch with NR waveforms.  We introduced the mismatch in \eqn{unfaithfulness} for calculating 
our calibration parameter, but now we shall use this as a test of the performance of \texttt{SEOBNR-GSF}.
We calculate the mismatch of \texttt{SEOBNR-GSF} for a set of \texttt{SXS} NR nonspinning waveforms, which were
also used in the calibration and assessment of \texttt{SEOBNRv5} in 
Refs.~\cite{Pompili:2023tna, vandeMeent:2023ols}.
The mismatches are calculated over a range of the binary's total mass range 
$10 M_{\odot} \leq M \leq 300 M_{\odot}$
for the $(\ell, m) = (2, 2)$ mode.

To assess the impact of adding the conservative 1GSF corrections to the Hamiltonian we compare to two other 
variants of the \texttt{SEOBNR} waveform family. We consider a version
of \texttt{SEOBNRv5} without any calibration of the Hamiltonian, only calibrating the $\Delta{t}_{\rm NR}$ 
parameter, which we will refer to as ``\texttt{SEOBNRv5} with $\Delta{t}_{\rm NR}$''. However, \texttt{SEOBNRv5} 
differs from \texttt{SEOBNR-GSF} not only in the physical content included in the Hamiltonian, but also in the 
choices of gauge and resummations. To get a more direct measure of the impact of including the conservative 1GSF 
corrections to the Hamiltonian, we also consider a version of \texttt{SEOBNR-GSF} in which we have turned these 
corrections off by setting $Z_{\rm fit}$ in \eqn{redshift_fit_form} to zero, replacing $Z_{\rm PN}$ by 
its $21.5$PN truncation, and re-calibrating $\Delta{t}_{\rm NR}$. 
This version will be labelled ``\texttt{SEOBNR-GSF-PN} with $\Delta{t}_{\rm NR}$" in the following.

\begin{figure}[h]
	\includegraphics[width=\linewidth]{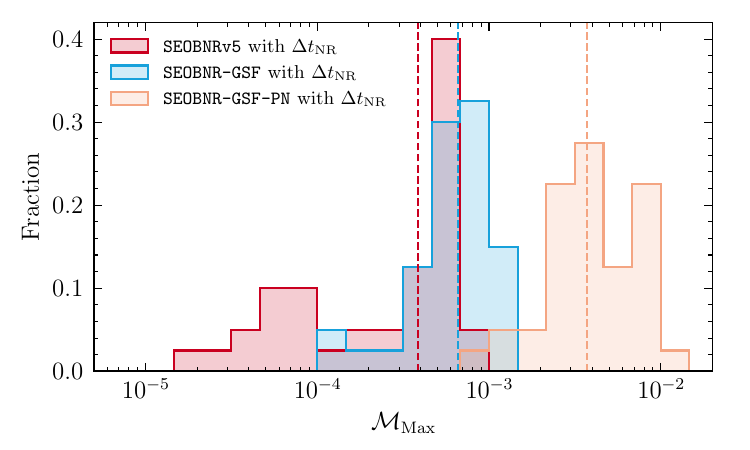}
	\caption{A plot of the distribution of the maximum mismatch, ${\cal M}_{\rm Max}$, for the calibrated models 
	we have presented in this work: \texttt{SEOBNR-GSF}, \texttt{SEOBNRv5} and \texttt{SEOBNR-GSF-PN}.
	This comparison was done with a
	public subset of nonspinning, circular \texttt{SXS} NR waveforms over the binary's total mass range from 
	$10 M_{\odot} \leq M \leq 300 M_{\odot}$.  The vertical dashed lines indicate the medians of the
	maximum mismatch.  Here we highlight the similar distributions and mismatch of the 
	\texttt{SEOBNR-GSF}, \texttt{SEOBNRv5}, but also emphasise the major improvement the inclusion the fit
	of GSF data has on the mismatch by comparing the distributions of \texttt{SEOBNR-GSF} and 
	\texttt{SEOBNR-GSF-PN}.
	}
	\label{fig:mismatch_cumulative}
\end{figure}

In \fig{mismatch_cumulative}, we plot the distribution of the maximum mismatch for the three models, with the 
medians of the mismatch distributions highlighted by the vertical dashed lines.
The median of \texttt{SEOBNR-GSF}, ${\cal M}_{\rm median} \sim 6.6 \times 10^{-4}$, falls just 
above that of \texttt{SEOBNRv5}, which in itself represents remarkably good agreement. 
\texttt{SEOBNR-GSF-PN}, on the other hand, has a median value of 
${\cal M}_{\rm median} \sim 3.7 \times 10^{-3}$, which is an order of magnitude worse than
\texttt{SEOBNRv5} and performs significantly worse than \texttt{SEOBNR-GSF} over the full range of
binary's total mass.
\fig{mismatch_cumulative} also emphasises the 
exceptional agreement of \texttt{SEOBNRv5} in the nonspinning case, which was also highlighted in 
Refs.~\cite{Pompili:2023tna, vandeMeent:2023ols}.  

\begin{figure}[h]
	\includegraphics[width=\linewidth]{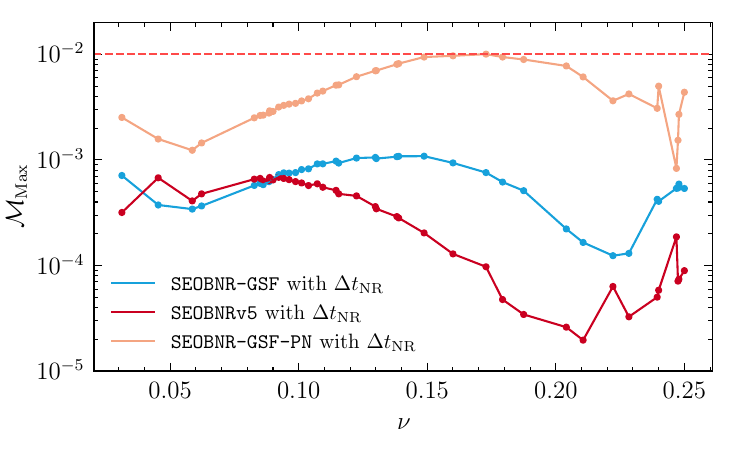}
	\caption{A comparison of maximum mismatch, ${\cal M}_{\rm Max}$, as a function of mass-ratio, $\nu$, 
	for the public subset of nonspinning, circular \texttt{SXS} NR waveforms, comparing the different calibrated models 
	we have presented in this work: \texttt{SEOBNR-GSF}, \texttt{SEOBNRv5} and \texttt{SEOBNR-GSF-PN}.
	For low mass-ratios, ${\cal M}_{\rm Max}$ of \texttt{SEOBNR-GSF} is comparable to that of \texttt{SEOBNRv5},
	but shows a decline in performance for more comparable masses. 
	Nevertheless, this underperformance remains significantly better than that of 
	\texttt{SEOBNR-GSF-PN}, where ${\cal M}_{\rm Max}$ increases dramatically with rising $\nu$, 
	underscoring the importance of including GSF data.
	}
	\label{fig:max_mismatch_comparison}
\end{figure}

The main question eminating from \fig{mismatch_cumulative} is whether the \texttt{SEOBNR-GSF} model is 
performing less well than \texttt{SEOBNRv5} broadly across the range of simulations, or whether there is
a specific region of parameter space for which \texttt{SEOBNR-GSF} underpeforms compared to \texttt{SEOBNRv5}.
We find that the maximum mismatch degrades for more comparable mass-ratios for \texttt{SEOBNR-GSF}.  
This is illustrated in \fig{max_mismatch_comparison}, which plots the maximum mismatch, ${\cal M}_{\rm Max}$, 
against the mass-ratio, $\nu$, for both models.
One observes that the improvement in maximum mismatch of \texttt{SEOBNR-GSF} relative to
\texttt{SEOBNR-GSF-PN} steadily grows with symmetric mass-ratio $\nu$, as one would expect from a 
higher order in $\nu$ correction.
On the other hand, the perfomance of \texttt{SEOBNR-GSF} relative to \texttt{SEOBNRv5}
is comparable for $\nu \lesssim 0.10$, but for increasing $\nu \rightarrow 0.25$
the difference between the mismatches of the two models worsens.  
In this region, the tuning of $\Delta t_{\rm NR}$ for \texttt{SEOBNR-GSF} degrades.
This was exemplified in \secref{delta_t_22_calibration}, as the calibration was hampered at these mass-ratios 
by issues of multimodality.

A key distinction between \texttt{SEOBNRv5} and \texttt{SEOBNR-GSF} lies in the inclusion of the Pad\'{e}-resummation 
of the Hamiltonian based on Taylor-expanded PN information. This resummation allows 
greater flexibility in the calibration  parameters, particularly $\Delta t_{\rm NR}$, which in turn improves 
agreement with NR. \texttt{SEOBNR-GSF} lacks such resummation in the Hamiltonian. 
Moreover, the PS gauge appears to be a suboptimal choice, despite resolving the LR 
divergence issue. 
This is evident in the EOB model dynamics, which show that in this specific gauge, the Hamiltonian remains too 
close to the probe limit ($\nu \rightarrow 0$) in the strong-field regime, as highlighted in 
Ref.~\cite{Antonelli:2019fmq}. Consequently, it lacks the flexibility afforded by the resummed \texttt{SEOBNRv5} 
Hamiltonian.  Our investigations further underscore the importance of resummation techniques, showing that without 
resumming the potentials in the \texttt{SEOBNRv5} Hamiltonian, the model’s 
posteriors are too weakly constrained to permit meaningful calibration of $\Delta t_{\rm NR}$.

\section{Summary and discussion}
\label{sec:summary_and_discussion}

In this paper we have constructed a full (quasicircular) inspiral-merger-ringdown EOB model, \texttt{SEOBNR-GSF}, 
that for its inspiral dynamics takes inputs only from GSF and PN results without any calibration to NR results 
in the EOB Hamiltonian and radiation-reaction force. Calibration to NR for this model happens 
for the merger and ringdown phase in the form of NQC corrections and calibration of the attachment time 
$\Delta t_{\rm NR}$. For the inspiral phase we include both 1GSF conservative corrections following 
Ref.~\cite{Antonelli:2019fmq} and 2GSF dissipative results following Ref.~\cite{vandeMeent:2023ols}. 
Consequently, the model should nominally achieve 1PA accuracy.

A central objective of this study was to assess the perfomance of \texttt{SEOBNR-GSF} model relative to alternative 
approaches, including the \texttt{SEOBNRv5} model and direct GSF calculations, particularly the \texttt{1PAT1} model. 
The most significant improvements in our model arise from the integration of GSF-based corrections, 
especially at lower mass ratios, where the dephasing between \texttt{SEOBNR-GSF} and NR waveforms remains small 
across the inspiral and early plunge phases. For low mass-ratio systems, \texttt{SEOBNR-GSF} exhibits phase 
differences of less than $0.1$ radians up to $900M$ before merger, which is comparable to 
\texttt{SEOBNRv5}.

An further noteworthy aspect of this study was the comparison between \texttt{SEOBNR-GSF} and direct GSF models, 
such as the \texttt{1PAT1} variant. 
As expected, \texttt{1PAT1} performs well in the early inspiral 
but begins to deviate substantially from NR as the binary approaches the ISCO. 
The multiscale expansion inherent in the \texttt{1PAT1} model leads to divergences that render it unreliable 
near ISCO, whereas the \texttt{SEOBNR-GSF} model, which does not rely on these approximations, continues to 
produce valid waveforms through the merger phase. This reinforces the benefit of adopting the EOB 
formalism, which naturally accommodates the entire coalescence process, as opposed to purely
GSF approach with a multiscale expansion of the \texttt{1PAT1} model, which is limited to the inspiral.

While \texttt{SEOBNR-GSF} remains reliable during the inspiral for a wide range of mass ratios, 
it performs less well during the late plunge and merger phases, particularly in the 
comparable-mass regime where $q \lesssim 1/5$, where phase discrepancies grow more pronounced. 
This degradation is largely due to the absence of a resummed Hamiltonian and the use of the PS-gauge, 
both of which limit the model’s fidelity in the strong-field regime. 
By contrast, \texttt{SEOBNRv5} benefits from more sophisticated resummation techniques that allow
the model to better capture the late-time dynamics. These differences highlight current limitations in 
the \texttt{SEOBNR-GSF} framework and motivate the need for further development.

A key feature of our approach was the minimal calibration effort required for \texttt{SEOBNR-GSF}. 
Unlike \texttt{SEOBNRv5}, which includes multiple calibration parameters to improve agreement with NR data, 
our model primarily relied on a single calibration parameter, $\Delta t_{\rm NR}$, which was tuned 
using NR data. While this simplified calibration procedure yielded good 
results in the low mass-ratio regime, it may limit the model’s flexibility when attempting to describe 
more complex systems, such as binaries with high-mass ratios or spinning components.

It is important to acknowledge that while \texttt{SEOBNR-GSF} shows promise, it does not yet match the overall 
accuracy of \texttt{SEOBNRv5}, particularly in the transition from inspiral to merger for higher mass ratios. 
One observation from our results is that the behaviour of \texttt{SEOBNR-GSF} near merger becomes increasingly 
unnatural for mass-ratios $q \lesssim 1/5$ leading to a rapid deterioration of the mismatch with NR.
The absence of a resummed Hamiltonian and the specific choice of PS gauge in \texttt{SEOBNR-GSF} 
lead to larger discrepancies in the strong-field regime. 
It will be important in follow-up work to examine alternative solutions.

Despite the limitations in terms of the choice of gauge and resummation, we show that the inclusion of the
strong-field numerical GSF data into the Hamiltonian can offer substantial improvements for full IMR EOB models.
We demonstrate this by comparing the \texttt{SEOBNR-GSF} model to a version that does not include the strong-field
data in the Hamiltonian, \texttt{SEOBNR-GSF-PN}.  Post-calibration, \texttt{SEOBNR-GSF} improves the maximum
mismatch by up-to an order of magnitude, strongly suggesting that further improvements may be possible with an 
alternative gauge choice.

There are several avenues for future improvement of the \texttt{SEOBNR-GSF} model. 
One of the most promising directions is the incorporation of spin effects, which are currently absent. 
Given the modular nature of the \texttt{pySEOBNR} framework, incorporating spin is a natural next step. 
Additionally, exploring alternative gauge choices and resummation techniques may further enhance 
the model’s performance, particularly in the comparable-mass regime.

\begin{acknowledgments}
The authors would like thank to Lorenzo Pompilli and Mohammed Khalil for their 
insightful discussions, which greatly contributed to the completion of this work. 
They also thank Raffi Enficiaud for his invaluable assistance with scientific computing.
The authors are also grateful to members of the Multiscale Self-Force (MSF) collaboration: 
Barry Wardell, Niels Warburton, and Adam Pound, for providing the new 2GSF data 
utilised in this work.
MvdM acknowledges financial support by 
the VILLUM Foundation (grant no. VIL37766), 
the DNRF Chair program (grant no. DNRF162) by the Danish National Research Foundation,
and the European Union’s Horizon ERC Synergy Grant “Making Sense of the Unexpected in the Gravitational-Wave Sky” grant agreement no.\ GWSky–101167314. 
Views and opinions expressed are however those of the authors only and do not necessarily reflect those of the European Union or the European Research Council. Neither the European Union nor the granting authority can be held responsible for them.
\end{acknowledgments}

\appendix

\section{Solutions to Hamilton's equations}
\label{sec:sols_hamiltons_eqs}
In this section we briefly reproduce the intermediate solutions of the two Hamilton's equations
used in the construction of the post-Schwarzschild gauge Hamiltonian potential, 
$\hat{Q}^{\rm PS}_{\rm SF}$-potential, in \secref{ps_gauge_hamiltonian}.
These solutions were previously reported by Antonelli \emph{et al.} in Ref.~\cite{Antonelli:2019fmq},
but are presented here for completeness.

First, the solution for the mass-reduced circular-orbit angular momentum, $\hat{p}^{\rm circ}_{\varphi}$,
to \eqn{pr_dot}, is given by
\begin{widetext}
\begin{align}
	\hat{p}_{\varphi}^{\rm circ}(u,\nu)=&\frac{1}{\sqrt{u(1-3u)}}
	+\nu\, \frac{(1 - 2u)^{2} }{4 (1 - 3u)^{3} \sqrt{u}}
	\bigg[2(1 - 2u)^{3}f_{0}(u) + 2(1 - 3u)^{3/2}f_{1} (u)
	+2(1 - 2u)(1 - 3u)f_{2}(u)\ln E^{-2}_{S} (u) \nn\\
	&-(1 - 2u)^{4}f^{\prime}_{0}(u) - (1 - 2u)(1 - 3u)^{3/2}f^{\prime}_{1}(u)
	-(1 - 2u)^{2}(1 - 3u)\ln E^{-2}_{S}(u)f^{\prime}_2(u)\bigg] 
	+ \mathcal{O}(\nu^{2}),
	\label{eq:p_phi_circ}
\end{align}
\end{widetext}
up to linear-order in $\nu$, where $f^{\prime}_{i}(u) := df_{i}/du$.

The other intermediate expression, which is a solution to \eqn{Omega} after inserting \eqn{p_phi_circ},
is an expression of the (circular) inverse radius, $u$, in terms of the gauge invariant radius, $x$.  
One finds up to linear-order in the mass-ratio,
\begin{widetext}
	\begin{align}
		u^{\text{circ}}(x,\nu) &= x + \frac{x\,\nu}{6 (1 - 3x)^{3/2}}	
		\bigg[ 4 - 20x + 24x^{2} - (4 - 12x) 
		\sqrt{1 - 3x} - 10(1-2 x)^{4} f_{0}(x) \nn\\
		&- 4\sqrt{1 - 3x} \left(1 - 5x + 6x^{2} \right) f_{1}(x)
		+ \big(4 - 28 x + 64 x^{2} - 48 x^{3} -(6 - 42x + 96x^{2} - 72x^{3})\ln E^{-2}(x)\big) f_{2}(x) \nn\\
		&+ \left(1 - 10x + 40x^{2} - 80x^{3} + 80x^{4} - 32x^{5}\right) 
		f^{\prime}_{0}(x)+\sqrt{1 - 3x}\left(1 - 7x + 16x^{2} - 12x^{3}\right) f^{\prime}_{1}(x)\nonumber\\
		&+ \left(1 - 9x + 30x^{2} - 44x^{3} + 24x^{4} \right)\ln E^{-2}_{\rm S}(x) f^{\prime}_{2}(x)
		\bigg] + \mathcal{O}(\nu^{2}).
		\label{eq:u_circ}
	\end{align}
\end{widetext}

\section{Mismatch Comparison}
\label{sec:mismatch_figures}
\begin{figure*}[t!]
\includegraphics[width=\linewidth]{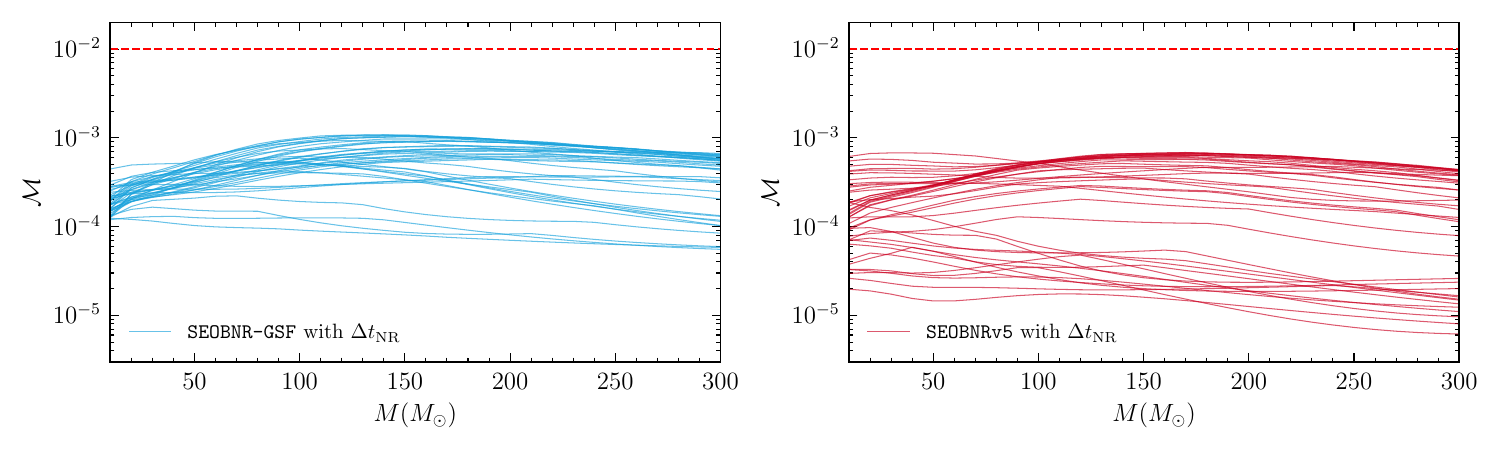}
\caption{Mismatches as a function of the total-mass of the binary, 
$10 M_{\odot} \leq M \leq 300 M_{\odot}$, for the two primary models featured in this work, 
\texttt{SEOBNRv5} and \texttt{SEOBNR-GSF}, calibrated with the $\Delta t_{\rm NR}$ parameter.
In this figure we use the complete set of nonspinning public \texttt{SXS} NR waveforms and we only
consider the $(\ell, m) = (2,2)$ mode.}
\label{fig:spaghetti_plot}
\end{figure*}

In this supplemtary appendix, we present \fig{spaghetti_plot}, 
a complimentary figure to \fig{max_mismatch_comparison}, that presents the mismatch
of the $(\ell, m) = (2,2)$ mode of \texttt{SEOBNRv5} and \texttt{SEOBNR-GSF} as a function 
of the binary's total mass ranging over $10 M_{\odot}$ to $300 M_{\odot}$.  
This is in contrast to \fig{mismatch_cumulative} and \fig{max_mismatch_comparison}, which only present the maximum mismatch across the same range of total
mass.  As previously, the models here are only calibrated through the $\Delta t_{\rm NR}$ parameter.

\section{Numerical-relativity simulations}
\label{sec:nr_sims}
In this work, we use a subset of publicly available numerical relativity (NR) simulations from the 
\texttt{SXS} collaboration~\cite{SXS:catalog,Boyle:2019kee} to calibrate and evaluate the performance of our EOB models. 
Table~\ref{tbl:sxs_sims} lists the simulations employed, along with key reference parameters. 
Specifically, we utilize nonspinning, quasicircular binaries across a range of mass ratios.
\input{SXSSims.tex}

%------------------------------------------------
%	Bibliography
%------------------------------------------------
\bibliography{Bibliography}
\end{document}

%% file: SXSSims.tex
{\renewcommand{\arraystretch}{1.4}
\begin{table*}[!tbp]
\begin{tabular*}{\textwidth}{l @{\extracolsep{\fill}} c c c r r c}
\hline
\hline
ID & \multicolumn{1}{c}{$q$} & \multicolumn{1}{c}{$\nu$} & \multicolumn{1}{c}{$e$} & \multicolumn{1}{c}{$\chi_{1}$} & \multicolumn{1}{c}{$\chi_{2}$} & Used in Fig.\\
\hline
\textbf{SXS:BBH:2325}               &    1.000   &    0.2500   &   $7.86 \times 10^{-5}$ &   $3.645 \times 10^{-5}$ &   $3.602 \times 10^{-5}$ &     \ref{fig:delta_t_22_fits}, \ref{fig:delta_omega_isco}, \ref{fig:binding_energy}, \ref{fig:waveform_comparison_low_q}, \ref{fig:mismatch_cumulative}, \ref{fig:max_mismatch_comparison}, \ref{fig:spaghetti_plot}\\
SXS:BBH:0198               &    1.202   &    0.2479   &   $2.04 \times 10^{-4}$ &   $-5.042 \times 10^{-5}$ &   $8.544 \times 10^{-5}$ &    \ref{fig:delta_omega_isco}, \ref{fig:mismatch_cumulative}, \ref{fig:max_mismatch_comparison}, \ref{fig:spaghetti_plot}\\
SXS:BBH:0310               &    1.221   &    0.2475   &   $7.88 \times 10^{-4}$ &   $1.456 \times 10^{-4}$ &   $9.706 \times 10^{-5}$ &   \ref{fig:delta_omega_isco}, \ref{fig:mismatch_cumulative}, \ref{fig:max_mismatch_comparison}, \ref{fig:spaghetti_plot}\\
SXS:BBH:1143               &    1.250   &    0.2469   &   $1.02 \times 10^{-4}$ &   $-1.365 \times 10^{-4}$ &   $-2.545 \times 10^{-5}$ &   \ref{fig:delta_omega_isco}, \ref{fig:mismatch_cumulative}, \ref{fig:max_mismatch_comparison}, \ref{fig:spaghetti_plot}\\
\textbf{SXS:BBH:2331}               &    1.500   &    0.2400   &   $5.77 \times 10^{-5}$ &   $-7.576 \times 10^{-5}$ &   $-6.795 \times 10^{-6}$ &   \ref{fig:delta_t_22_fits}, \ref{fig:delta_omega_isco}, \ref{fig:mismatch_cumulative}, \ref{fig:max_mismatch_comparison}, \ref{fig:spaghetti_plot}\\
SXS:BBH:0194               &    1.517   &    0.2394   &   $8.02 \times 10^{-4}$ &   $3.193 \times 10^{-5}$ &   $-8.569 \times 10^{-5}$ &   \ref{fig:delta_omega_isco}, \ref{fig:mismatch_cumulative}, \ref{fig:max_mismatch_comparison}, \ref{fig:spaghetti_plot}\\
SXS:BBH:1354               &    1.832   &    0.2284   &   $4.79 \times 10^{-5}$ &   $-1.500 \times 10^{-4}$ &   $1.264 \times 10^{-4}$ &   \ref{fig:delta_omega_isco}, \ref{fig:mismatch_cumulative}, \ref{fig:max_mismatch_comparison}, \ref{fig:spaghetti_plot}\\
\textbf{SXS:BBH:2425}              &    2.000   &    0.2222   &   $2.89 \times 10^{-4}$ &   $-7.663 \times 10^{-5}$ &   $1.160 \times 10^{-4}$ &   \ref{fig:multimodal_distributions}, \ref{fig:delta_t_22_fits}, \ref{fig:delta_omega_isco}, \ref{fig:mismatch_cumulative}, \ref{fig:max_mismatch_comparison}, \ref{fig:spaghetti_plot}\\
\textbf{SXS:BBH:0201}               &    2.316   &    0.2106   &   $1.41 \times 10^{-4}$ &   $6.266 \times 10^{-5}$ &   $-4.156 \times 10^{-5}$ &   \ref{fig:delta_omega_isco}, \ref{fig:mismatch_cumulative}, \ref{fig:max_mismatch_comparison}, \ref{fig:spaghetti_plot}\\
\textbf{SXS:BBH:0259}               &    2.450   &    0.2041   &   $4.90 \times 10^{-4}$ &   $9.373 \times 10^{-8}$ &   $2.478 \times 10^{-7}$ &   \ref{fig:delta_t_22_fits}, \ref{fig:delta_omega_isco}, \ref{fig:mismatch_cumulative}, \ref{fig:max_mismatch_comparison}, \ref{fig:spaghetti_plot}\\
\textbf{BFI:RerunCatalog\_SHK:003}   &    3.000   &    0.1875   &   $9.64 \times 10^{-5}$ &   $4.358 \times 10^{-6}$ &   $3.132 \times 10^{-6}$ &   \ref{fig:delta_t_22_fits}, \ref{fig:delta_omega_isco}, \ref{fig:binding_energy}, \ref{fig:mismatch_cumulative}, \ref{fig:max_mismatch_comparison}, \ref{fig:spaghetti_plot}\\
SXS:BBH:0200               &    3.272   &    0.1792   &   $4.14 \times 10^{-4}$ &   $-5.023 \times 10^{-5}$ &   $-1.095 \times 10^{-5}$ &   \ref{fig:delta_omega_isco}, \ref{fig:mismatch_cumulative}, \ref{fig:max_mismatch_comparison}, \ref{fig:spaghetti_plot}\\
\textbf{SXS:BBH:0294}               &    3.500   &    0.1728   &   $4.34 \times 10^{-5}$ &   $-2.312 \times 10^{-7}$ &   $3.533 \times 10^{-7}$ &   \ref{fig:delta_t_22_fits}, \ref{fig:delta_omega_isco}, \ref{fig:mismatch_cumulative}, \ref{fig:max_mismatch_comparison}, \ref{fig:spaghetti_plot}\\
SXS:BBH:2483               &    3.500   &    0.1728   &   $2.67 \times 10^{-4}$ &   $-2.705 \times 10^{-5}$ &   $6.294 \times 10^{-5}$ &   \ref{fig:delta_omega_isco}, \ref{fig:mismatch_cumulative}, \ref{fig:max_mismatch_comparison}, \ref{fig:spaghetti_plot}\\
SXS:BBH:2499               &    4.000   &    0.1600   &   $1.57 \times 10^{-4}$ &   $8.406 \times 10^{-5}$ &   $3.421 \times 10^{-6}$ &   \ref{fig:delta_omega_isco}, \ref{fig:waveform_comparison_low_q}, \ref{fig:mismatch_cumulative}, \ref{fig:max_mismatch_comparison}, \ref{fig:spaghetti_plot}\\
\textbf{BFI:RerunCatalog\_SHK:004}   &    4.000   &    0.1600   &   $1.57 \times 10^{-4}$ &   $8.406 \times 10^{-6}$ &   $3.420 \times 10^{-6}$ &   \ref{fig:delta_t_22_fits}, \ref{fig:delta_omega_isco}, \ref{fig:mismatch_cumulative}, \ref{fig:max_mismatch_comparison}, \ref{fig:spaghetti_plot}\\
\textbf{BFI:q8\_7d:0527}             &    4.500   &    0.1488   &   $2.36 \times 10^{-4}$ &   $4.219 \times 10^{-5}$ &   $2.865 \times 10^{-5}$ &   \ref{fig:delta_t_22_fits}, \ref{fig:delta_omega_isco}, \ref{fig:mismatch_cumulative}, \ref{fig:max_mismatch_comparison}, \ref{fig:spaghetti_plot}\\
\textbf{SXS:BBH:2374}               &    5.000   &    0.1389   &   $5.30 \times 10^{-4}$ &   $-8.130 \times 10^{-5}$ &   $5.244 \times 10^{-5}$ &   \ref{fig:delta_t_22_fits}, \ref{fig:delta_omega_isco}, \ref{fig:mismatch_cumulative}, \ref{fig:max_mismatch_comparison}, \ref{fig:spaghetti_plot}\\
SXS:BBH:0187               &    5.039   &    0.1382   &   $4.60 \times 10^{-5}$ &   $8.798 \times 10^{-6}$ &   $-1.204 \times 10^{-5}$ &   \ref{fig:delta_omega_isco}, \ref{fig:binding_energy}, \ref{fig:mismatch_cumulative}, \ref{fig:max_mismatch_comparison}, \ref{fig:spaghetti_plot}\\
\textbf{SXS:BBH:0296}               &    5.500   &    0.1302   &   $3.30 \times 10^{-5}$ &   $2.513 \times 10^{-7}$ &   $3.072 \times 10^{-7}$ &   \ref{fig:delta_t_22_fits}, \ref{fig:delta_omega_isco}, \ref{fig:mismatch_cumulative}, \ref{fig:max_mismatch_comparison}, \ref{fig:spaghetti_plot}\\
SXS:BBH:2486               &    5.500   &    0.1302   &   $4.44 \times 10^{-4}$ &   $2.801 \times 10^{-6}$ &   $-9.807 \times 10^{-6}$ &   \ref{fig:delta_omega_isco}, \ref{fig:mismatch_cumulative}, \ref{fig:max_mismatch_comparison}, \ref{fig:spaghetti_plot}\\
SXS:BBH:0197               &    5.522   &    0.1298   &   $2.20 \times 10^{-4}$ &   $-3.702 \times 10^{-5}$ &   $-1.524 \times 10^{-5}$ &   \ref{fig:delta_omega_isco}, \ref{fig:mismatch_cumulative}, \ref{fig:max_mismatch_comparison}, \ref{fig:spaghetti_plot}\\
\textbf{SXS:BBH:2164}               &    5.999   &    0.1225   &   $3.85 \times 10^{-4}$ &   $-2,709 \times 10^{-6}$ &   $-1.417 \times 10^{-5}$ &   \ref{fig:delta_t_22_fits}, \ref{fig:delta_omega_isco}, \ref{fig:mismatch_cumulative}, \ref{fig:max_mismatch_comparison}, \ref{fig:spaghetti_plot}\\
\textbf{SXS:BBH:0297}               &    6.500   &    0.1155   &   $5.29 \times 10^{-5}$ &   $6.329 \times 10^{-7}$ &   $5.524 \times 10^{-7}$ &   \ref{fig:delta_t_22_fits}, \ref{fig:delta_omega_isco}, \ref{fig:mismatch_cumulative}, \ref{fig:max_mismatch_comparison}, \ref{fig:spaghetti_plot}\\
SXS:BBH:2488               &    6.500   &    0.1155   &   $7.26 \times 10^{-4}$ &   $2.792 \times 10^{-5}$ &   $-2.407 \times 10^{-5}$ &   \ref{fig:delta_omega_isco}, \ref{fig:mismatch_cumulative}, \ref{fig:max_mismatch_comparison}, \ref{fig:spaghetti_plot}\\
SXS:BBH:0912               &    6.578   &    0.1145   &   $5.02 \times 10^{-5}$ &   $2.510 \times 10^{-5}$ &   $-5.071 \times 10^{-5}$ &   \ref{fig:delta_omega_isco}, \ref{fig:mismatch_cumulative}, \ref{fig:max_mismatch_comparison}, \ref{fig:spaghetti_plot}\\
\textbf{SXS:BBH:0298}              &    7.000   &    0.1094   &   $4.00 \times 10^{-5}$ &   $6.904 \times 10^{-7}$ &   $5.823 \times 10^{-7}$ &   \ref{fig:delta_t_22_fits}, \ref{fig:delta_omega_isco}, \ref{fig:mismatch_cumulative}, \ref{fig:max_mismatch_comparison}, \ref{fig:spaghetti_plot}\\
SXS:BBH:2491               &    7.000   &    0.1094   &   $3.61 \times 10^{-4}$ &   $1.137 \times 10^{-5}$ &   $4.506 \times 10^{-5}$ &   \ref{fig:delta_omega_isco}, \ref{fig:mismatch_cumulative}, \ref{fig:max_mismatch_comparison}, \ref{fig:spaghetti_plot}\\
SXS:BBH:0188               &    7.187   &    0.1072   &   $1.61 \times 10^{-4}$ &   $1.523 \times 10^{-6}$ &   $-2.445 \times 10^{-5}$ &   \ref{fig:delta_omega_isco}, \ref{fig:mismatch_cumulative}, \ref{fig:max_mismatch_comparison}, \ref{fig:spaghetti_plot}\\
\textbf{SXS:BBH:0299}               &    7.500   &    0.1038   &   $5.60 \times 10^{-5}$ &   $7.131 \times 10^{-7}$ &   $5.960 \times 10^{-7}$ &   \ref{fig:delta_t_22_fits}, \ref{fig:delta_omega_isco}, \ref{fig:mismatch_cumulative}, \ref{fig:max_mismatch_comparison}, \ref{fig:spaghetti_plot}\\
SXS:BBH:2490               &    7.500   &    0.1038   &   $5.52 \times 10^{-4}$ &   $-2.915 \times 10^{-5}$ &   $-5.938 \times 10^{-6}$ &   \ref{fig:delta_omega_isco}, \ref{fig:mismatch_cumulative}, \ref{fig:max_mismatch_comparison}, \ref{fig:spaghetti_plot}\\
SXS:BBH:0195               &    7.761   &    0.1011   &   $2.24 \times 10^{-4}$ &   $1.319 \times 10^{-5}$ &   $-4.008 \times 10^{-5}$ &   \ref{fig:delta_omega_isco}, \ref{fig:mismatch_cumulative}, \ref{fig:max_mismatch_comparison}, \ref{fig:spaghetti_plot}\\
\textbf{BFI:q8\_7d:0086}             &    8.000   &   0.09876   &   $3.41 \times 10^{-4}$ &   $-5.603 \times 10^{-5}$ &   $6.588 \times 10^{-5}$ &   \ref{fig:delta_t_22_fits}, \ref{fig:delta_omega_isco}, \ref{fig:mismatch_cumulative}, \ref{fig:max_mismatch_comparison}, \ref{fig:spaghetti_plot}\\
SXS:BBH:0186               &    8.267   &   0.09626   &   $6.70 \times 10^{-4}$ &   $1.017 \times 10^{-6}$ &   $-8.824 \times 10^{-8}$ &   \ref{fig:delta_omega_isco}, \ref{fig:mismatch_cumulative}, \ref{fig:max_mismatch_comparison}, \ref{fig:spaghetti_plot}\\
\textbf{SXS:BBH:0300}               &    8.500   &    0.0942   &   $6.00 \times 10^{-5}$ &   $8.720 \times 10^{-7}$ &   $7.241 \times 10^{-7}$ &   \ref{fig:delta_t_22_fits}, \ref{fig:delta_omega_isco}, \ref{fig:mismatch_cumulative}, \ref{fig:max_mismatch_comparison}, \ref{fig:spaghetti_plot}\\
SXS:BBH:2492               &    8.501   &   0.09417   &   $8.58 \times 10^{-4}$ &   $-3.198 \times 10^{-6}$ &   $-1.826 \times 10^{-5}$ &   \ref{fig:delta_omega_isco}, \ref{fig:mismatch_cumulative}, \ref{fig:max_mismatch_comparison}, \ref{fig:spaghetti_plot}\\
SXS:BBH:0199               &    8.729   &   0.09222   &   $6.77 \times 10^{-5}$ &   $-1.113 \times 10^{-6}$ &   $-3.313 \times 10^{-5}$ &   \ref{fig:delta_omega_isco}, \ref{fig:mismatch_cumulative}, \ref{fig:max_mismatch_comparison}, \ref{fig:spaghetti_plot}\\
\textbf{SXS:BBH:0301}               &    9.000   &    0.0900   &   $5.70 \times 10^{-4}$ &   $8.874 \times 10^{-7}$ &   $7.440 \times 10^{-7}$ &   \ref{fig:delta_t_22_fits}, \ref{fig:delta_omega_isco}, \ref{fig:mismatch_cumulative}, \ref{fig:max_mismatch_comparison}, \ref{fig:spaghetti_plot}\\
SXS:BBH:2495               &    9.001   &   0.08999   &   $2.01 \times 10^{-4}$ &   $1.365 \times 10^{-6}$ &   $-8.766 \times 10^{-6}$ &   \ref{fig:delta_omega_isco}, \ref{fig:mismatch_cumulative}, \ref{fig:max_mismatch_comparison}, \ref{fig:spaghetti_plot}\\
SXS:BBH:0189               &    9.167   &   0.08868   &   $8.17 \times 10^{-5}$ &   $1.183 \times 10^{-5}$ &   $-6.788 \times 10^{-6}$ &   \ref{fig:delta_omega_isco}, \ref{fig:mismatch_cumulative}, \ref{fig:max_mismatch_comparison}, \ref{fig:spaghetti_plot}\\
SXS:BBH:1108               &    9.200   &   0.08843   &   $1.48 \times 10^{-4}$ &   $-2.253 \times 10^{-6}$ &   $-1.464 \times 10^{-6}$ &   \ref{fig:delta_omega_isco}, \ref{fig:mismatch_cumulative}, \ref{fig:max_mismatch_comparison}, \ref{fig:spaghetti_plot}\\
SXS:BBH:2494               &    9.497   &   0.08619   &   $1.59 \times 10^{-4}$ &   $-1.568 \times 10^{-5}$ &   $-3.539 \times 10^{-5}$ &   \ref{fig:delta_omega_isco}, \ref{fig:mismatch_cumulative}, \ref{fig:max_mismatch_comparison}, \ref{fig:spaghetti_plot}\\
\textbf{SXS:BBH:0302}               &    9.500   &   0.08617   &   $5.40 \times 10^{-4}$ &   $9.056 \times 10^{-7}$ &   $7.641 \times 10^{-7}$ &   \ref{fig:delta_t_22_fits}, \ref{fig:delta_omega_isco}, \ref{fig:mismatch_cumulative}, \ref{fig:max_mismatch_comparison}, \ref{fig:spaghetti_plot}\\
SXS:BBH:0196               &    9.663   &   0.08499   &   $2.63 \times 10^{-4}$ &   $1.673 \times 10^{-6}$ &   $-2.731 \times 10^{-5}$ &   \ref{fig:delta_omega_isco}, \ref{fig:mismatch_cumulative}, \ref{fig:max_mismatch_comparison}, \ref{fig:spaghetti_plot}\\
\textbf{SXS:BBH:0185}               &    9.990   &   0.08272   &   $2.93 \times 10^{-4}$ &   $1.281 \times 10^{-5}$ &   $-1.434 \times 10^{-5}$ &   \ref{fig:delta_t_22_fits}, \ref{fig:delta_omega_isco}, \ref{fig:binding_energy}, \ref{fig:waveform_comparison_high_q}, \ref{fig:mismatch_cumulative}, \ref{fig:max_mismatch_comparison}, \ref{fig:spaghetti_plot}\\
\textbf{SXS:BBH:2480}               &    14.00   &   0.06222   &   $3.81 \times 10^{-4}$ &   $7.625 \times 10^{-6}$ &   $-4.136 \times 10^{-6}$ &   \ref{fig:delta_t_22_fits}, \ref{fig:delta_omega_isco}, \ref{fig:waveform_comparison_high_q}, \ref{fig:mismatch_cumulative}, \ref{fig:max_mismatch_comparison}, \ref{fig:spaghetti_plot}\\
\textbf{SXS:BBH:2477}               &    15.00   &   0.05859   &   $3.69 \times 10^{-4}$ &   $6.433 \times 10^{-6}$ &   $-4.521 \times 10^{-6}$ &   \ref{fig:delta_t_22_fits}, \ref{fig:delta_omega_isco}, \ref{fig:mismatch_cumulative}, \ref{fig:max_mismatch_comparison}, \ref{fig:spaghetti_plot}\\
\textbf{SXS:BBH:2325}               &    20.00   &   0.04536   &   $2.51 \times 10^{-4}$ &   $3.434 \times 10^{-5}$ &   $-1.022 \times 10^{-4}$ &   \ref{fig:delta_t_22_fits}, \ref{fig:delta_omega_isco}, \ref{fig:mismatch_cumulative}, \ref{fig:max_mismatch_comparison}, \ref{fig:spaghetti_plot}\\
\hline

\hline
\hline
\end{tabular*}
\caption{A table summarising the numerical relativity simulations from the SXS collaboration used in this work.
The initial eccentricity is denoted by $e$, while $\chi_{1}$ and $\chi_{2}$ are the reference spins of the component black holes.
The final column indicates the figures in which each simulation appears. Simulations used in the calibration procedure are indicated in \textbf{bold}.}
\label{tbl:sxs_sims}
\end{table*}
}